\documentclass[review]{elsarticle}

\usepackage{graphicx}
\usepackage{float}
\usepackage{amsmath}
\usepackage{gensymb}
\usepackage[caption=false]{subfig}
\usepackage{caption} 
\usepackage[margin=1.35in]{geometry}
\usepackage{array}
\newcolumntype{P}[1]{>{\centering\arraybackslash}m{1in}}

\usepackage[utf8]{inputenc}

\journal{Journal of Nuclear Instruments and Methods in Physics Research Section A}

\begin{document}
	
	\begin{frontmatter}
		
	\title{Design, construction, and characterization of a compact DD neutron generator designed for $^{40}$Ar/$^{39}$Ar geochronology}
	\author[UCB,lbl]{Mauricio Ayllon\corref{mycorrespondingauthor}}
	\ead{mayllon@berkeley.edu}
	\author[UCB]{Parker A. Adams}
	\author[livermore]{Joseph D. Bauer}
    \author[UCB]{Jon C. Batchelder}
	\author[geochronology]{Tim A. Becker}
	\author[UCB,lbl]{Lee A. Bernstein}
	\author[UTK]{Su-Ann Chong}
	\author[UCB]{Jay James}
	\author[UCB,livermore]{Leo E. Kirsch}
	\author[UCB]{Ka-Ngo Leung}
	\author[UCB]{Eric F. Matthews}
	\author[UCB]{Jonathan T. Morrell}
	\author[geochronology]{Paul R. Renne}
	\author[uml]{Andrew M. Rogers}
	\author[geochronology]{Daniel Rutte}
	\author[UCB,lbl]{Andrew S. Voyles}
	\author[UCB]{Karl Van Bibber}
	\author[livermore]{Cory S. Waltz}

	\cortext[mycorrespondingauthor]{Corresponding author}
	
	\address[UCB]{Department of Nuclear Engineering, University of California, Berkeley, CA 94720, USA}
	\address[lbl]{Lawrence Berkeley National Laboratory, Berkeley, CA 94720, USA}
	\address[livermore]{Lawrence Livermore National Laboratory, Livermore, CA 94550, USA}
	\address[geochronology]{Berkeley Geochronology Center, 2455 Ridge Road,Berkeley, CA 94709, USA}
	\address[uml]{University of Massachusetts Lowell, 220 Pawtucket St, Lowell, MA 01854, USA}
	\address[UTK]{University of Tennessee, Knoxville, TN 37996, USA}

	
\begin{abstract}
	
A next-generation, high-flux DD neutron generator has been designed, commissioned, and characterized, and is now operational in a new facility at the University of California Berkeley. The generator, originally designed for $^{40}$Ar/$^{39}$Ar dating of geological materials, has since served numerous additional applications, including medical isotope production studies, with others planned for the near future. In this work, we present an overview of the High Flux Neutron Generator (HFNG) which includes a variety of simulations, analytical models, and experimental validation of results. Extensive analysis was performed in order to characterize the neutron yield, flux, and energy distribution at specific locations where samples may be loaded for irradiation. A notable design feature of the HFNG is the possibility for sample irradiation internal to the cathode, just 8 mm away from the neutron production site, thus maximizing the neutron flux (n/cm$^2$/s). The generator's maximum neutron flux at this irradiation position is 2.58 $\times$10$^7$ n/cm$^2$/s $\pm$ 5\% (approximately 3$\times$10$^8$ n/s total yield) as measured via activation of small natural indium foils. However, future development is aimed at achieving an order of magnitude increase in flux. Additionally, the deuterium ion beam optics were optimized by simulations for various extraction configurations in order to achieve a uniform neutron flux distribution and an acceptable heat load. Finally, experiments were performed in order to benchmark the modeling and characterization of the HFNG.   
		
\end{abstract}
		
	
\end{frontmatter}
	
	
\section{Introduction}

Fusion-based neutron generators are used in many fields of research, education, and industry. As opposed to many other neutron sources such as research reactors, neutron generators provide nearly monoenergetic neutrons, pose a lower proliferation risk, do not produce high-level waste, and can be made compact and portable. The most common reactions for such generators are shown in Equations \ref{eq:reactionDD} (``DD'') and \ref{eq:reactionDT} (``DT''). At low interaction energies, DT reactions benefit from a resonance near 100 keV, which allows for a higher neutron yield at a given beam current. Around this energy, the neutron yield from a DT source is about two orders of magnitude higher than that of a DD source. However, the DT neutron energy spectrum is also much higher i.e. around 14.1 MeV, which can be a significant disadvantage, or even inapplicable for some purposes, such as the production of $^{39}$Ar for dating of geological samples, as explained later on. Moreover, tritium poses greater radiological handling and licensing challenges. 

\begin{align} \label{eq:reactionDD}
&^{2}H + ^{2}H \rightarrow ^{3}He + n + 3.27\ MeV\ (50\%) \\
&^{2}H + ^{3}H \rightarrow ^{4}He + n + 17.6\ MeV \label{eq:reactionDT}
\end{align}

Neutron generators are commonly used for a variety of applications including activation analysis, radioisotope production, fast neutron imaging, oil well logging, fundamental nuclear research, and nuclear data measurements. The High Flux Neutron Generator (HFNG) located at the University of California, Berkeley was designed and built by the Department of Nuclear Engineering and the Berkeley Geochronology Center (BGC), with its primary motivation being the irradiation of samples for the $^{40}$Ar/$^{39}$Ar dating method. $^{40}$Ar/$^{39}$Ar dating is based on the natural decay of $^{40}$K to $^{40}$Ar with the daughter accumulating in a natural terrestrial or planetary rock, mineral, or an archaeological artifact over time. The sample is irradiated with neutrons transmuting $^{39}$K via the reaction $^{39}$K(n,p)$^{39}$Ar with the $^{39}$Ar acting as a tracer of potassium content thereafter. The production of $^{39}$Ar is determined by co-irradiation of a standard of known geological age. Argon is released from the sample in vacuo and analyzed with a mass spectrometer, so a geological age can be calculated.
Conventionally, $^{40}$Ar/$^{39}$Ar samples are irradiated with fission spectrum neutrons in research reactors. The wide spectrum of neutron energies poses two major drawbacks to the method. (i) Reactions on K, Ca, Cl and Br (e.g., $^{42}$Ca(n,$\alpha$)$^{39}$Ar) produce interfering isotopes that need to be corrected for, increasing the uncertainty on calculated ages significantly \cite{turner1971}, \cite{RUTTE20171}. (ii) Kinetic energy of the neutrons is transferred to the produced $^{39}$Ar; proton emission and decay from its high-energy states result in partial loss of the $^{39}$Ar from the material \cite{turner2}. This recoil effect is governed by the volume to surface ratio of the sample and poses a lower limit on the reliably dateable grain size. Utilization of DD neutrons can ameliorate both drawbacks: The near monoenergetic neutrons with energy around 2.7 MeV eliminates or reduces the undesired interference reactions and is anticipated to reduce recoil effects significantly \cite{geochron}.
 
Since its commissioning, the HFNG has been proven useful for a variety of additional applications including Prompt Gamma Neutron Activation Analysis (PGNAA) for on-site gamma detector calibration, cross-section measurements for emerging medical isotopes \cite{np_paper}, the study of delayed gamma rays from fission of $^{238}$U, single event upset (SEU) of CPUs to develop radiation hardened electronics, the study of NaI detector response in neutron fields, and most recently, cross-section measurements of $^{35}$Cl(n,p)$^{35}$S and $^{35}$Cl(n,$\alpha$)$^{32}$P, which are of significant interest for the design of molten salt reactors. 

The HFNG operates by accelerating positive deuterium ions towards a titanium target electrode biased up to -120 kV. These ions embed in the target matrix forming titanium deuteride (T$_i$D$_x$) where $x$ ranges from 1 to 2, forming an implanted target for subsequent deuterium ions incident upon the cathode to initiate the DD fusion reaction shown in Equation \ref{eq:reactionDD}. Neutrons are born with a well characterized distribution of energies and relative yields with respect to the angle formed with the incident beam (taken to be $0^{\circ}$). Precise characterization of the neutron spectrum throughout the HFNG was crucial for several of the applications previously outlined, requiring that simulation tools and experimental validation be developed together.
	
\section{Description of the Facility and the Neutron Generator}
	
\begin{figure}
	\centering
	\includegraphics[width=1\textwidth]{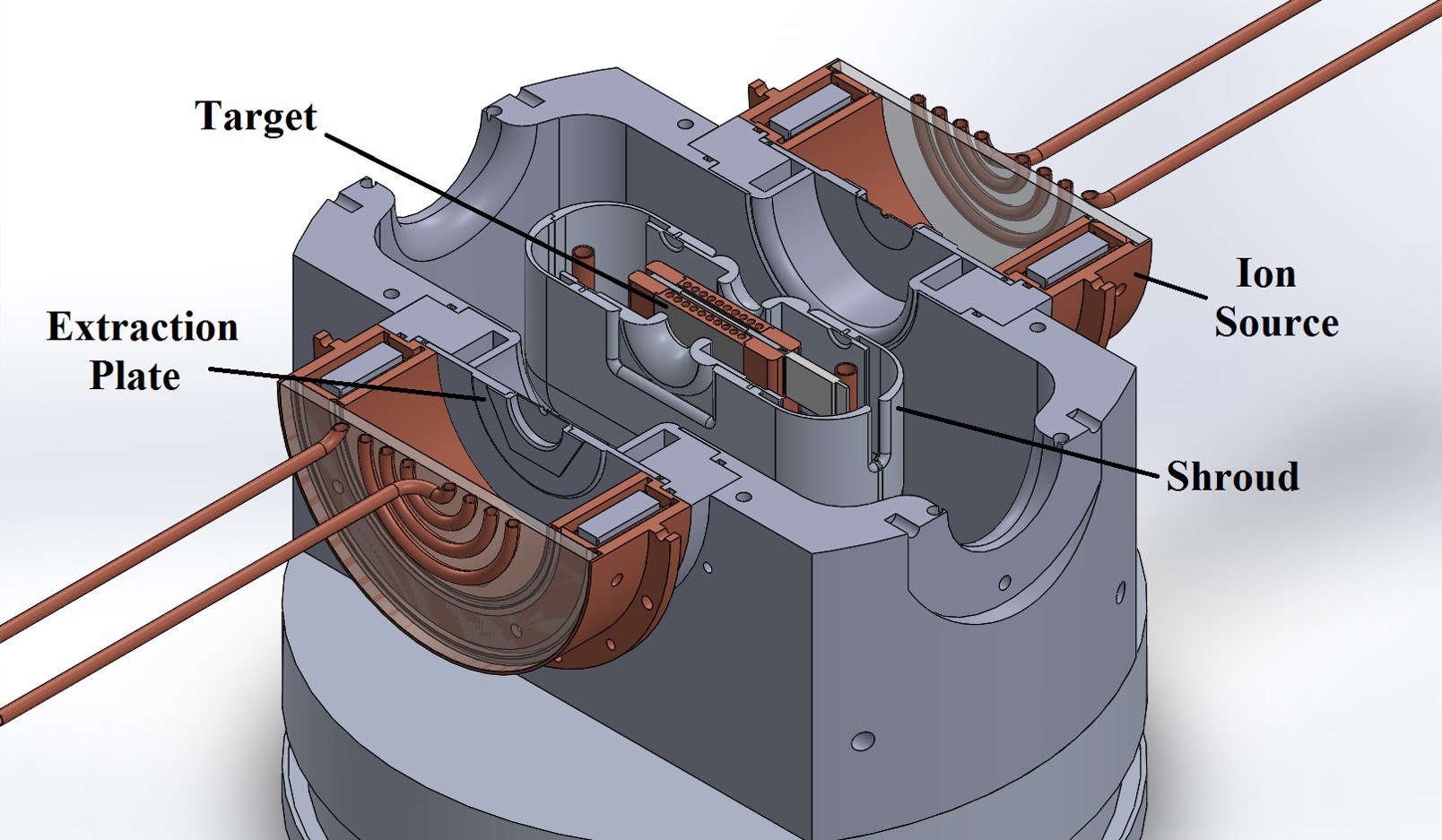}
	\caption{Cross-sectional view of the HFNG exposing the shroud, target, extraction plate (plasma electrode), and ion sources.}
	\label{fig:HFNG}
\end{figure}

The HFNG is designed around two radio frequency (RF)-driven multicusp ion sources which straddle a titanium-coated copper target, as shown in Figure \ref{fig:HFNG}. This arrangement allows for doubling the deuterium current incident upon the target, which effectively doubles the neutron flux at the sample location. This type of ion source, based upon similar designs developed at Lawrence Berkeley National Lab \cite{RFsource}, produces positive deuterium ions by ionizing deuterium gas with an RF field of 13.56 MHz. The ions are confined in copper chambers with quartz windows. The neodymium magnets surrounding the copper chambers allow for operation of the ion sources at lower pressures since their purpose is to confine the electrons for longer periods of time hence allowing for more ions produced per electron \cite{WuThesis}. The quartz windows are used in order to increase the proportion of monatomic deuterium ions because the recombination rate for heavier molecular species in insulators is much lower than on metal surfaces. Therefore, the major benefit of this ion source design is the production of predominantly monatomic deuterium ions \cite{WuThesis}, which is important since diatomic and triatomic deuterium ions achieve lower ultimate energies at specified extraction potentials. Because the DD fusion cross-section is a monotonically increasing function of the incident deuteron energy up to about 2.2 MeV (the deuteron breakup energy), lower bias voltages reduce the neutron yield.   

\begin{figure}
	\centering
	\includegraphics[width=1\textwidth]{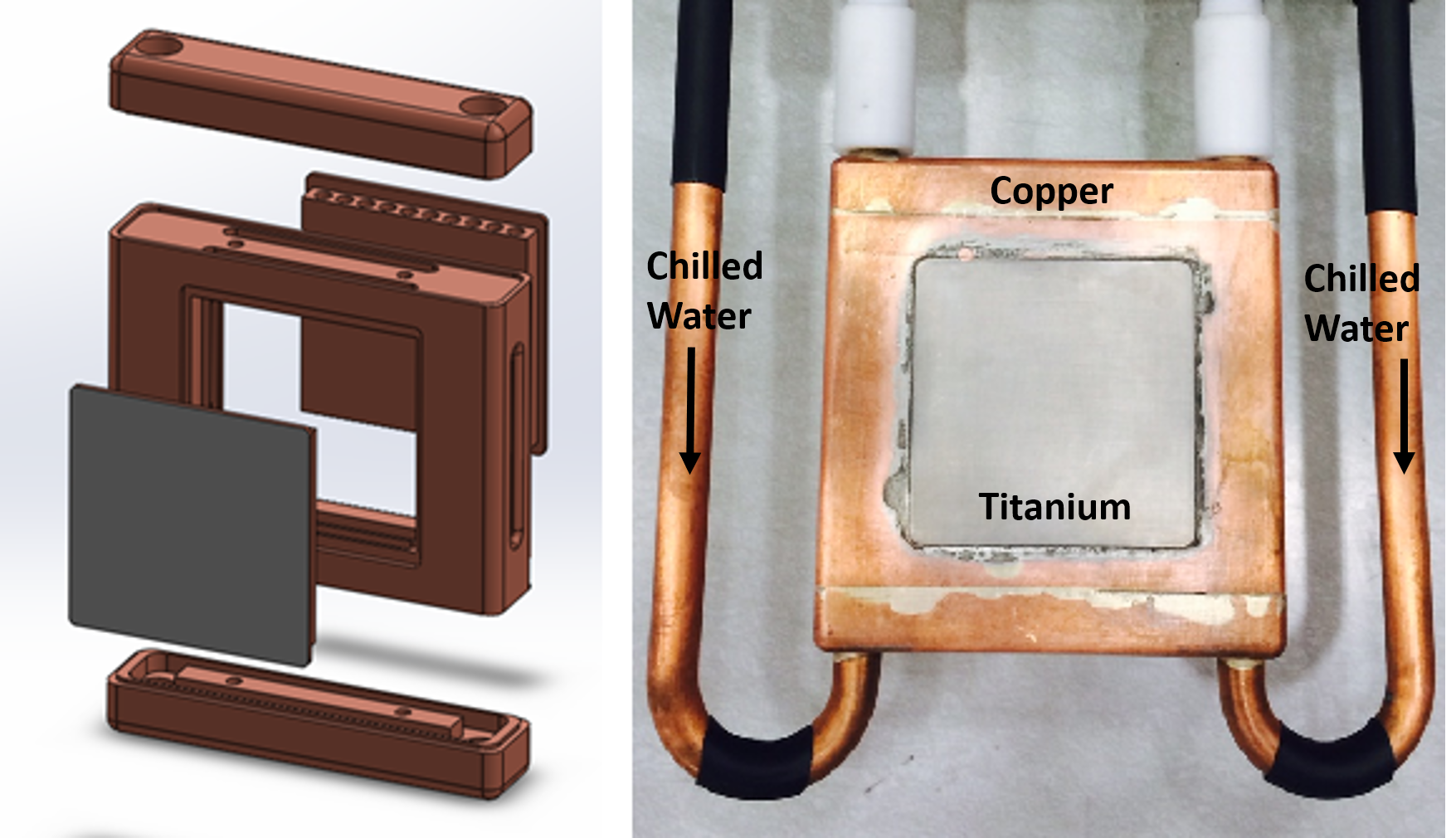}
	\caption{Target assembly showing the CAD design of the faceplate and the actual target soldered in place. The gray titanium-coated copper production surface is visible in the center of the target, and cooling lines for deionized water can be seen on the sides of the target.}
	\label{fig:new_target}
\end{figure}

The target, shown in Figure \ref{fig:new_target}, is primarily composed of copper due to its excellent thermal conductivity. However, copper does not form hydrides as well as titanium\cite{CRC}, so a thin layer (120 microns) of titanium is either explosion-bonded or diffusion-bonded (both kinds of targets were used) onto the copper structure to enhance deuteron implantation and retention. The unique design of this target allows for samples to be placed very close to the neutron production surface ($\approx 8$ mm separation), hence maximizing the neutron flux at the sample location. Active internal cooling with deionized water is necessary in order to dissipate the heat load generated by the deuterium beam incident upon the surface of the target. This is particularly important given the fact that deuterium diffuses out of titanium at temperatures above 200$^{\circ}$C \cite{CRC}. The target is biased at a negative potential (up to -120 kV) in order to extract the positive deuterium ions.

\begin{figure}
	\centering
	\includegraphics[width=1\textwidth]{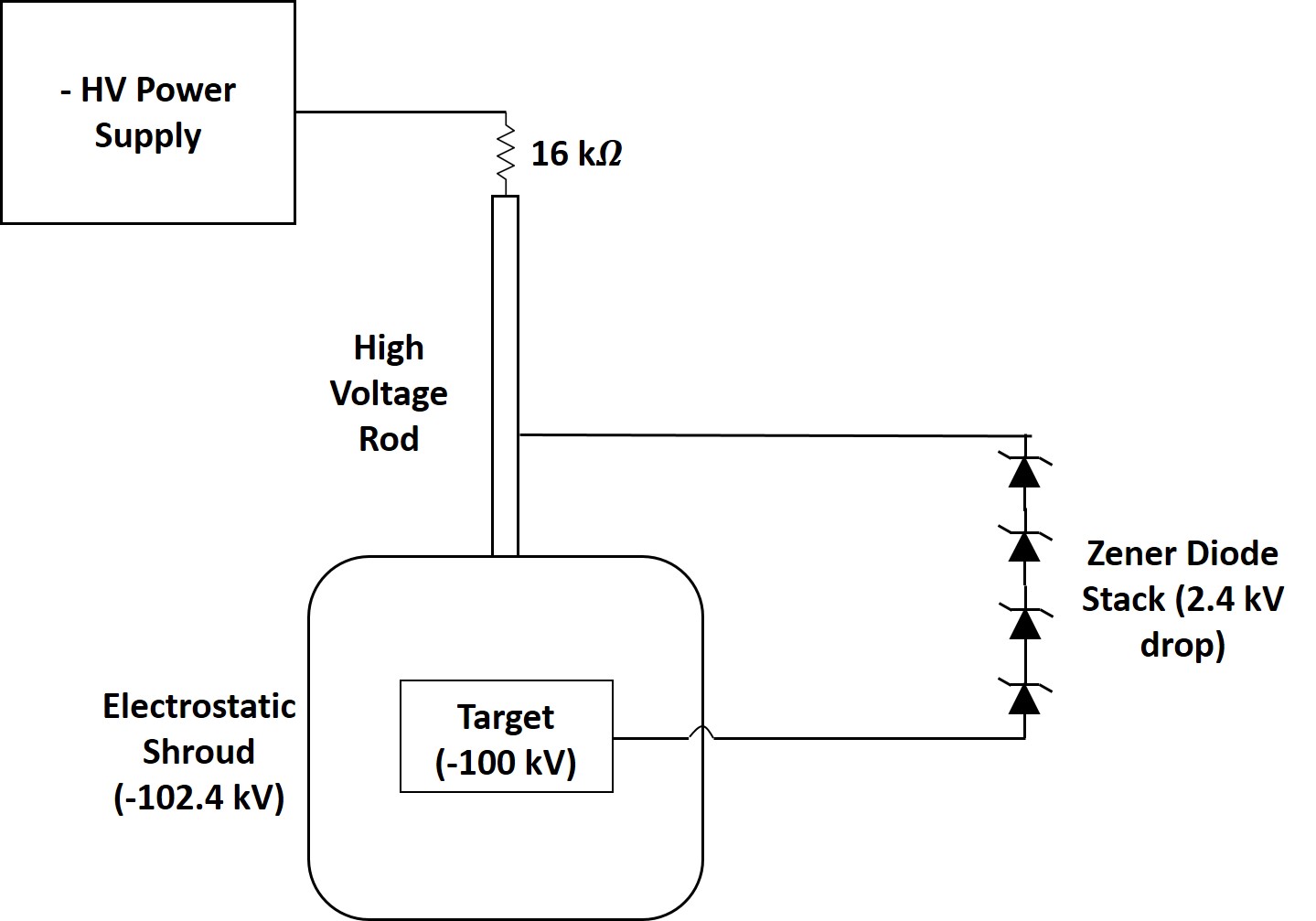}
	\caption{Circuit schematic of the diode assembly for typical operation at -100 kV target bias potential. Note that the voltage differential between the shroud and the target can be varied from 0 to 2.4 kV.}
	\label{fig:diodes}
\end{figure}

The target is encased in an electrostatic shroud biased at a more negative potential (up to $\Delta V=2.4$ kV) through an arrangement of Zener diodes in order to reverse the direction of the electric field inside this structure. The circuit diagram is shown in Figure \ref{fig:diodes}. Without a shroud, electrons that are sputtered off the titanium target by the deuterium ions would experience a repulsive electrostatic force, accelerating them back towards the plasma electrode to cause arcing, overheating (or even melting) of surrounding structures, and a very large flux of bremsstrahlung X-rays. The electric field reversal inside the shroud makes the electrons experience an attractive force which guides them back towards the target, hence mitigating the issues mentioned above. A detailed description of this technique has already been published \cite{electronSup}. 

The HFNG's internal vacuum chamber is evacuated by a scroll pump connected in series with a turbo-pump. The ultimate backing pressure is approximately $2\times10^{-6}$ Torr, while the operating pressure is in the lower $10^{-5}$ Torr when employing both ion sources. The scroll pump is oil-free, since the HFNG produces a certain amount of tritium during operation, which readily binds to oil molecules. Furthermore, a scroll pump can be operated continuously without the risk of backstreaming fluids.
	
A polyethylene structure was constructed that can be assembled around the HFNG to thermalize neutrons for different types of experiments. In general, nuclear reactions of the type $(n,\gamma)$ have higher absorption cross-sections at lower energies.

There is also an external beam port aligned with the neutron emitting spot of the HFNG in order to perform experiments outside the vault. Some experiments that have been performed using the external beam line include characterization of gamma detectors in a neutron field, and prompt gamma activation analysis of various materials.   
	
The HFNG vacuum line includes a residual gas analyzer (RGA) system that allows for continuous, on-line monitoring of the gas composition inside the chamber. It serves as both a leak detector, as well as a reliable indicator of water vapor, which causes a variety of problems during operation such as a noticeable increase in arcing.
The HFNG components and surrounding structures were carefully chosen so as to avoid long-lived neutron activation products. For instance, the vacuum chamber and most of the structural components are made of aluminum, which presents only one relevant activation channel i.e. 27Al(n,g)28Al \cite{NDS28} which decays by beta minus (t$_{1/2}=$2.245 min) to an excited state of $^{28}$Si which in turn emits a 1.78 MeV gamma ray. Because of this short half-life, HFNG personnel can enter the vault safely after around 20 min following irradiation. It was calculated that the structural components with greatest activation levels are the copper ion sources with a saturation dose rate at a few centimeters away from them of 3 mRem/h assuming a neutron source of 3$\times$10$^{10}$ n/s (an order of magnitude higher than the current configuration). Additionally, the vault is constantly monitored for gamma radiation levels with an ion chamber and periodic swipes are performed for tritium monitoring with a liquid scintillator counter (LSC).

\section{Modeling of the Neutron Yield, Flux, and Energy Distribution}

The absolute neutron yield can be indirectly inferred from both flux measurements and the corresponding neutron source modeling, which is done using the Monte Carlo code MCNP6 \cite{MCNP}. Moreover, we also present a semi-analytical approach based on the interaction processes within the target which can be used in order to gain greater insight and guide future neutron source developments. Additionally, it is important to model the neutron energy distribution in the lab frame for a variety of experiments including cross-section measurements \cite{np_paper}.
	
\subsection{Neutron Yield Analysis}

A fraction of the accelerated deuterons do not undergo DD fusion at the energy specified by the bias voltage due to two main processes: 1) deuterons slow down as they penetrate into the titanium target before a fusion reaction occurs, and 2) the deuterium ion beam is not purely monatomic, which also lowers the interaction energy. Both of these processes result in a lower neutron yield since the DD fusion cross-section decreases with decreasing deuteron ion energy. The deuterium ions slow down mainly because of collisions with electrons in the target. This process can be described using data for the energy loss per unit distance (stopping power) of deuterons in titanium \cite{SRIM}. Additionally, there are monatomic (D$^+$), diatomic (D$_2^+$), and triatomic (D$_3^+$) ion species in the deuterium plasma. For heavier species, the final energy they achieve for a given target potential is lower. For instance, molecular (diatomic) deuterium accelerated to 100 keV will share this energy equally between the two deuterium nuclei, \emph{i.e.}, each ion has an energy of 50 keV each. For RF ion sources, the deuterium ions are predominantly monatomic (over 90\%, in some cases \cite{WuThesis}). However, the exact composition is a function of the RF power, the ion source operating pressure, the structural materials, and the geometry \cite{CRC}, \cite{surfRecomb}. The HFNG ion sources have not yet been characterized in terms of ion species. This process requires mass spectrometric analysis. 

\begin{equation} \label{eq:yield}
Y(E_d)=\phi_d n_d \int_{0}^{E_{max}} \frac{\sigma(E_d)}{dE/dx(E_d)}dE_d
\end{equation}

The neutron yield can be approximated based on an analytical approach outlined in \cite{CRC} together with empirical data for the stopping power ($dE/dx$) of deuterons in titanium, the cross-section for the neutron-producing reaction in Equation \ref{eq:reactionDD}, the fraction of different deuterium species in the ion source, and the deuterium-to-titanium ratio in the self-loaded target when saturated (TiD$_2$). Equation \ref{eq:yield} was used to estimate the neutron yield as a function of deuteron energy, where $\phi_d$ is the deuteron flux (cm$^{-2}$s$^{-1}$) incident on the target (ion current per unit charge), $n_d$ is the number density of deuterons in the target (cm$^{-3}$) when saturated, $\sigma(E_d)$ (cm$^2$) \cite{Cross_sections} is the cross section for the DD neutron-producing reaction, and $dE/dx(E_d)$ \cite{SRIM} is the energy loss per unit distance in titanium. Note that the lower limit of integration is taken to be zero since the Q-value of the reaction is positive. The yield estimated with this equation can be compared to experiments coupled with the MCNP6 model. The shortcomings of this approximation is our lack of knowledge regarding the exact ion fraction of the beam and the saturation conditions of the target, as explained before.

\subsection{Neutron Flux and Energy Distribution}

\begin{figure}
	\centering
	\includegraphics[width=1\textwidth]{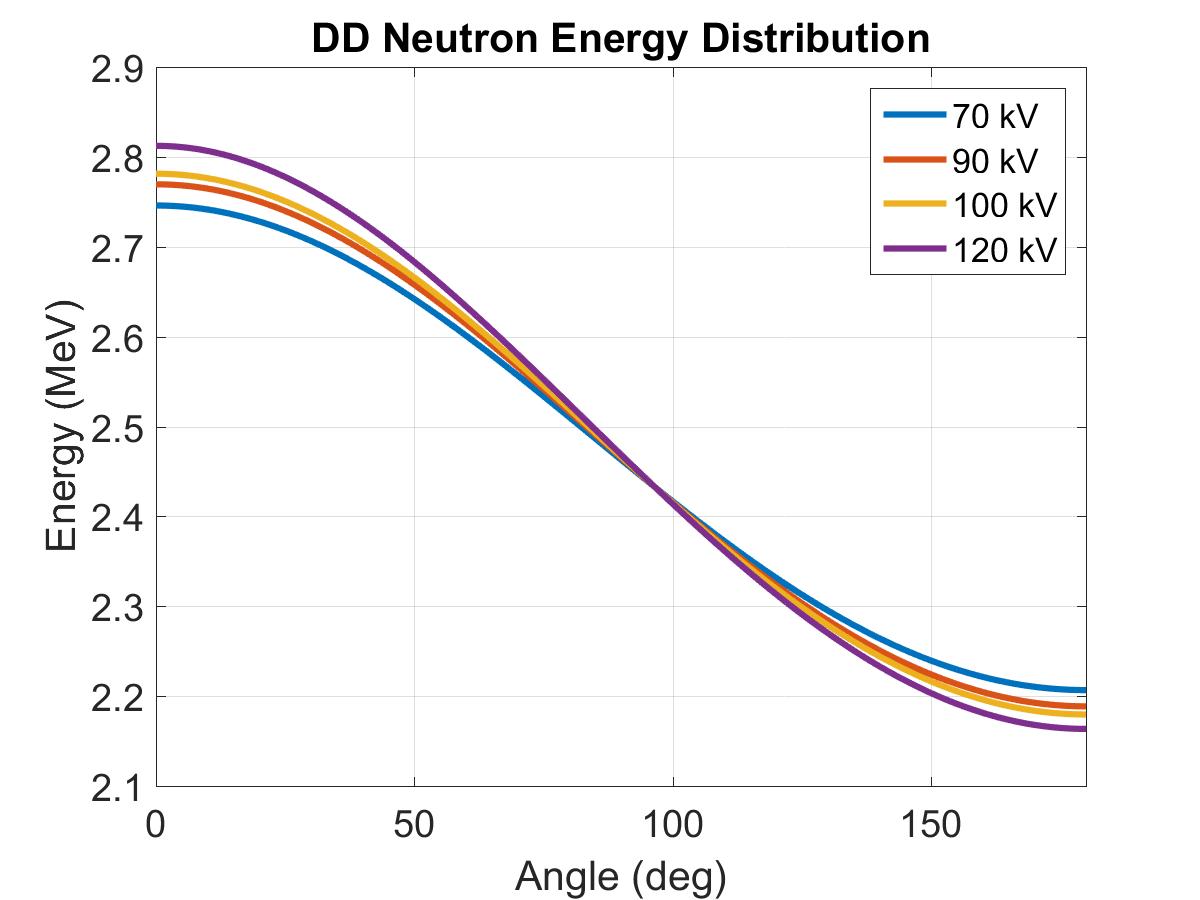}
	\caption{Neutron energy in the lab system as a function of angle. With increasing ion energy, a wider spread in neutron energy is seen over the 0$^{\circ}$ - 180$^{\circ}$ range \cite{Cross_sections}.}
	\label{fig:energy_dist}
\end{figure}

The kinematics of the 2-body interaction between deuterons results in a well-characterized variation of neutron energy as a function of angle in the lab frame \cite{CRC}. This neutron energy-angle variation is seen for a range of ion energies in Figure \ref{fig:energy_dist}, clearly displaying that this variation is quite significant. Additionally, the energy-angle distribution changes as a function of incident deuteron energy. For instance, at zero deuteron incident kinetic energy, the emitted neutron would have an energy of 2.45 MeV, while at 100 keV, the maximum is close to 2.8 MeV. This seeming paradox is accounted for by the redistribution of the total available energy for the reaction (Q-value plus incident deuteron energy) between the $^3$He nucleus and the neutron. 

\begin{figure}
	\centering
	\includegraphics[width=0.7\textwidth]{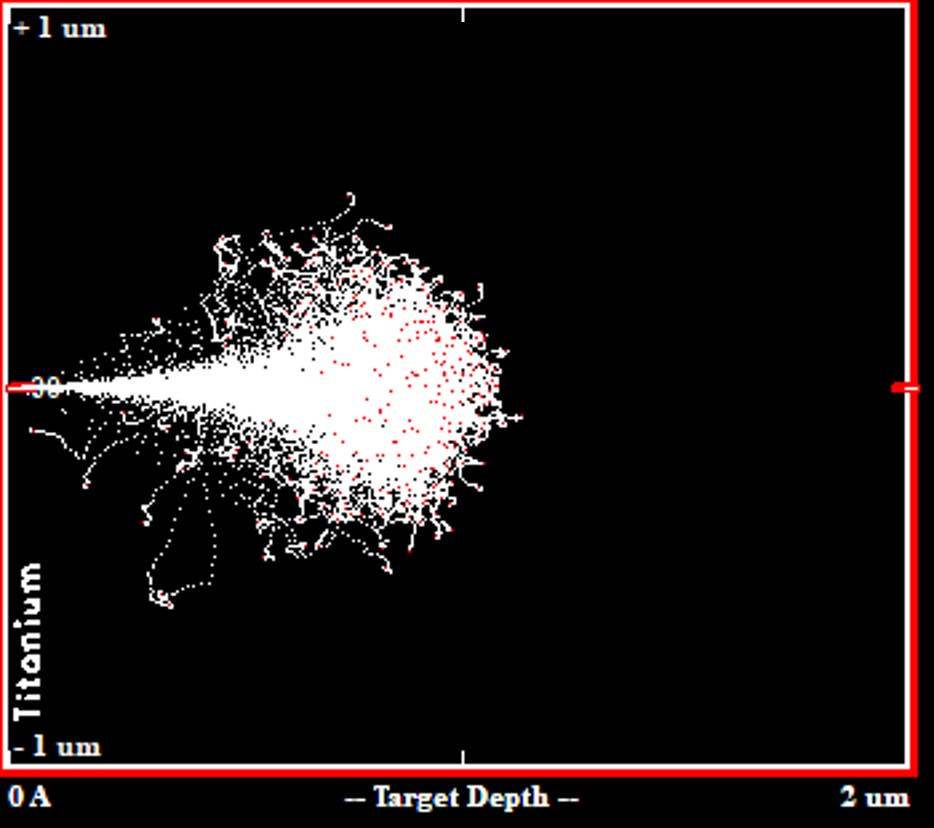}
	\caption{TRIM simulation of 120 keV D$^+$ ions on titanium. The longitudinal range is around 0.85 microns \cite{SRIM}.}
	\label{fig:srim}
\end{figure}

The energy spectrum broadens as the neutron travels through the target, leading to different empirical correlations for thin and thick targets \cite{CRC}. A target is considered thick if all the accelerated deuterons either interact or stop within the target. Since the range of 120 keV monatomic deuterons in titanium is slightly below 1 micron, as shown in Figure \ref{fig:srim}, the HFNG target is considered to be thick.

The angular distribution is well-fitted with an expansion in Legendre polynomials, as shown in Equation \ref{eqn:energDist}. Note that this equation is valid for DD and DT reactions below 500 keV deuteron energy for both, thin and thick targets. 

\begin{equation}
E_n(E_d,\theta)=E_0+\sum_{i=1}^{n}E_i cos^i \theta
\label{eqn:energDist}
\end{equation}

The coefficients $E_i$ were determined for a few energy values of deuteron energy \cite{CRC}. Therefore, they were interpolated in our model using a cubic spline to be applicable for a wider range of desired energy.

Furthermore, the neutron yield is also dependent on the emitted angle, and DD neutrons exhibit a larger anisotropy than DT neutrons \cite{CRC}. The yield in this design of DD neutron generators is forward-focused, with the maximum yield in the direction of the incident deuteron beam ($0^\circ$). The relative yield, defined as the total yield normalized to that at $90^{\circ}$ (lowest yield), can be described by Equation~\ref{eqn:angDist}, a best-fit model of the available data \cite{Cross_sections}.

\begin{equation}
\frac{Y(\theta)}{Y(90^{\circ})}=A_0+\sum_{i=1}^{n}A_i cos^i \theta
\label{eqn:angDist}
\end{equation}

\begin{figure}[H]
	\begin{center}
		\includegraphics[width=1\textwidth]{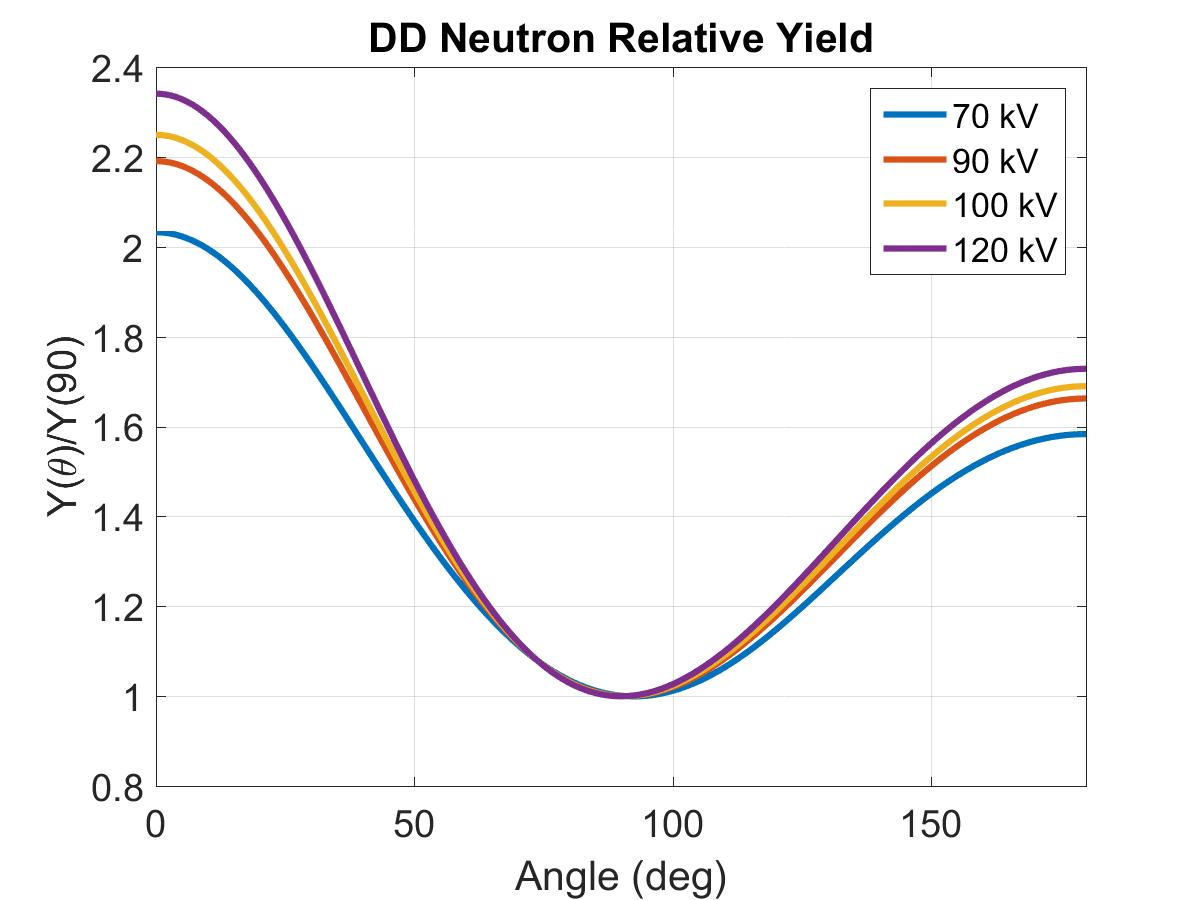} 
		\caption{Relative neutron yield for different accelerating voltages. The neutron yield becomes increasingly more pronounced with larger voltage at forward and backward angles.}
		\label{fig:yield}
	\end{center}
\end{figure}

The coefficients $A_i$ were interpolated using a cubic spline for various deuteron energies, with the resulting relative yield distributions in the lab frame shown in Figure~\ref{fig:yield}.

\begin{figure}
	\centering
	\includegraphics[width=1\textwidth]{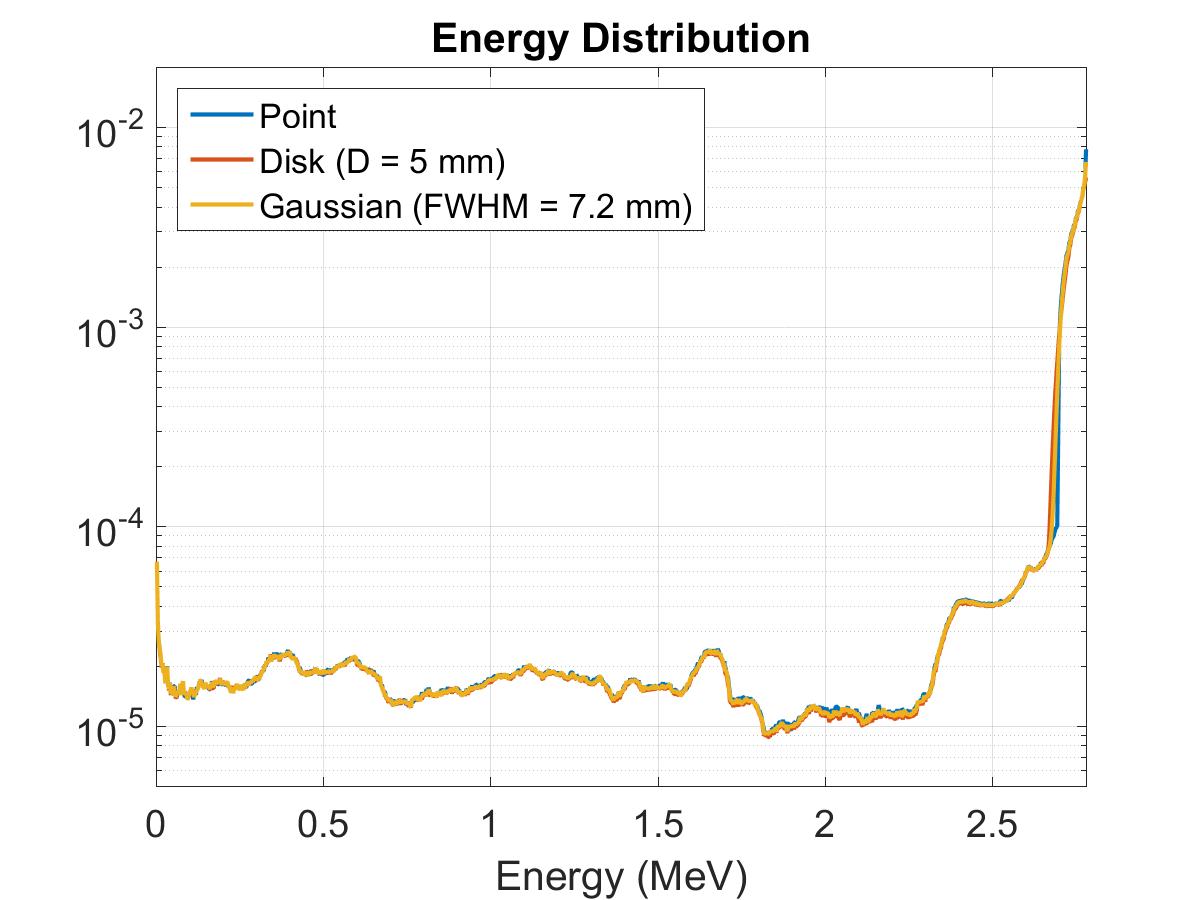}
	\caption{MCNP6-simulated energy distributions at the center of the sample holder location for different source definitions. The sample is taken to be 10 mm in diameter. Flux units on the y-axis are normalized per source neutron (MCNP6 F4 tally default units).}
	\label{fig:source_definitions}
\end{figure}

An MCNP6 \cite{MCNP} model of the generator and surrounding structures was developed, with the neutron source modeled according to experimental data obtained from \cite{Cross_sections}. The primary purpose of such a model is to obtain the flux and energy distributions at different irradiation locations. Different types of neutron sources were considered as inputs for the MCNP6 simulations based on experimental observations and ion beam optics simulations, which show that for small extraction apertures, the resulting beam profile on the surface of the target is more uniform (disk shaped), while for larger apertures, the beam profile is more Gaussian-shaped. Further details about this are presented in the following section. As a result, the types of neutron sources considered in the model were point-like, disk (D = 5 mm), and Gaussian shaped (FWHM = 7.2 mm ), with the latter taken from a best-fit model to ion beam optics simulations data. The results shown in this section apply only to a single-hole plasma electrode plate of 0.262 cm in diameter, which was used for several experiments including the one detailed in \cite{np_paper}. The code developed allows for different input parameters to be modified according to the experiment being performed. Such parameters include the acceleration voltage, the deuterium ion current, the atomic deuterium ion fraction in the plasma, the beam diameter, and the Gaussian profile of the beam. These results are shown in Figure \ref{fig:source_definitions}. Note that there is no significant difference among the three source definitions. However, if the irradiation location or the size of the samples change, it is important to re-do the simulations. Also note that less than 1\% of the neutrons incident on the sample can be attributed to scatter in the target or surrounding structures.  

\begin{figure}
	\centering
	\includegraphics[width=1\textwidth]{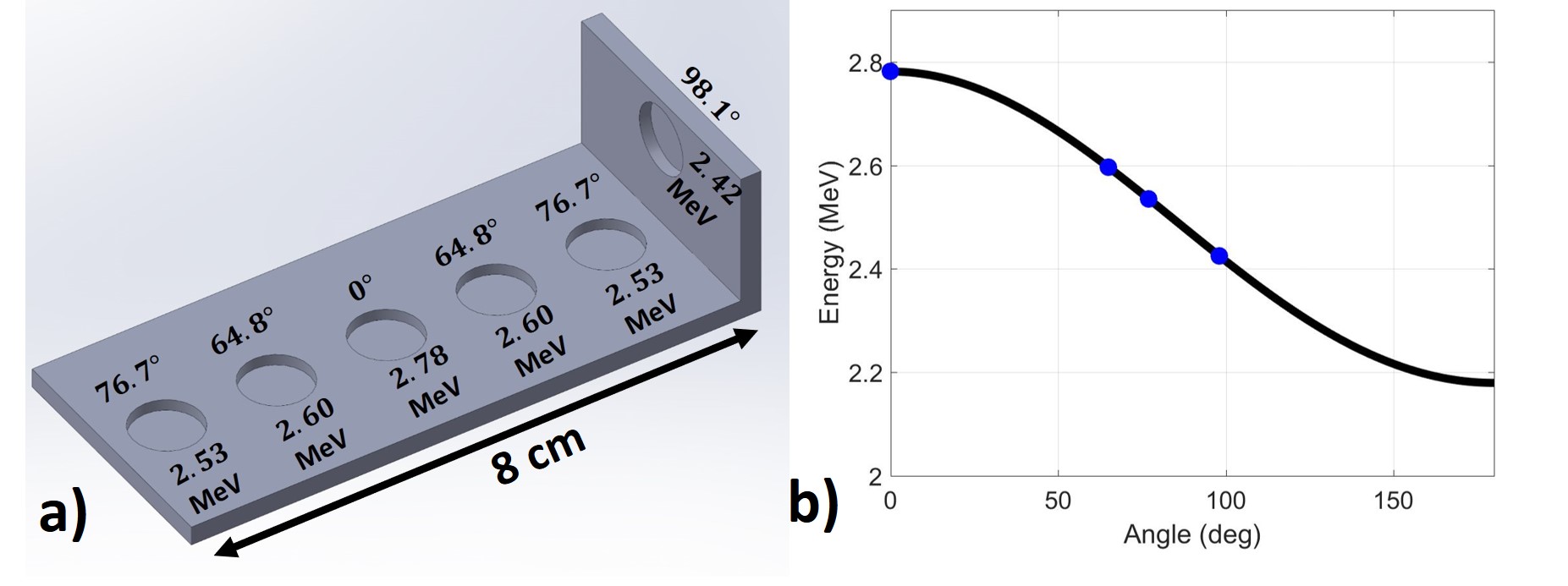}
	\caption{Sample holder that allows for irradiation of samples at different neutron energies. The energy values shown correspond to the center of each slot at 100 kV extraction voltage. The sample slots diameter is 10 mm each.}
	\label{fig:L_holder}
\end{figure}

In order to take advantage of the variation of neutron energy as a function of angle, an L-shaped sample holder, shown in Figure \ref{fig:L_holder} was designed so that multiple samples can be loaded to span neutron energies between 2.4 - 2.8 MeV at 100 kV extraction voltage. This arrangement is particularly interesting for cross section measurements, as it permits measurements at different neutron energies in a single irradiation. It has been used for $^{35}$Cl(n,p) and $^{35}$Cl(n,$\alpha$) reaction cross section measurements. The energy window depends on the size of the sample, and MCNP simulations should be performed in order to quantify this energy spread. The sample slot located at 98.1$^{\circ}$ allows for a ``clean'' neutron beam, which means that there is virtually no structural material in between the sample and the neutron production surface. This arrangement allows for a more narrow energy distribution since no scattering occurs on structural materials, but at the expense of a reduced neutron flux due to the $1/r^2$ dependence and the kinematics of the reaction yield, as shown in Figure \ref{fig:yield}. 

\section{Ion Beam Extraction Analysis}

The deuterium ion beam is extracted from the plasma electrode of the ion source through either a single or multiple apertures. The configuration and design of such an extraction mechanism largely determines the beam profile and heat deposition on the target. Precise knowledge and control of the ion optics is necessary in order to 1) achieve a uniform beam profile on the surface of the target and 2) prevent the localized temperature from exceeding $\approx 200^{\circ}\ C$ \cite{AANG}. The first requirement ensures that the resulting neutron flux is also uniform over the sample to be irradiated, which is essential for experiments such as the irradiation of geological samples \cite{geochron}. The second requirement has to do with the fact that implanted deuterium diffuses out (degases) from the titanium surface around this temperature \cite{CRC}, which in turn, results in a lower neutron output. Moreover, if the heat load surpasses a certain value determined by the specific target configuration, the target can be eroded and destroyed. It becomes increasingly difficult to meet the temperature requirements already at approximately 1 kW/cm$^2$, even though that estimate varies depending on the cooling capacity of the target and the beam spot size.

The optimum beam spot diameter varies according to the application of the neutron generator. For example, a very small spot size (approximately 1-2 mm) is needed for fast neutron imaging because it enhances the sharpness of the image \cite{adams}. However, this requirement limits the beam current the target can handle, which in turn, limits the neutron yield at a given acceleration voltage. The HFNG is designed to operate with a variety of exchangeable plasma electrode designs, which result in different beam profiles depending on the application. Extensive analysis and modeling were performed for a single extraction aperture and a multiple-aperture plasma electrodes. 

A flat plasma meniscus at extraction is desired in order to achieve a uniform beam profile \cite{CoryThesis}. However, precise modeling of the plasma meniscus is complicated due to all the variables involved and the imperfect fidelity of plasma physics simulations (especially near plasma boundaries). Moreover, slight changes in electron temperature and ion density in the plasma significantly affect the shape of the plasma meniscus, as determined by Bohm's Equation \cite{Plasma}. As the diameter of the extraction aperture decreases, the effects of the plasma meniscus shape on the resulting beam profile are greatly reduced, to the point that, at very small ($\sim$ 1 mm) apertures, the extracted beam is barely affected by focusing or defocusing effects due to the shape of the plasma meniscus. Additionally, the beam spreading along the acceleration gap is more pronounced with small-diameter apertures due to the shape of the equipotential lines near and around the aperture, as explained in the following sub-sections. This effect can be beneficial for limiting the heat load on the target. Therefore, it can be more advantageous to extract ions through several small apertures rather than through one equivalent aperture when designing high-flux neutron generators.      

The finite-element software package COMSOL Multiphysics 5.2b \cite{comsol} was used to simulate the ion beam trajectories for a single extraction aperture of 0.262 cm diameter and a 19-hole plasma electrode (each 1 mm inner diameter), as shown in Figure \ref{fig:single_vs_mult}. The geometry of the setup is accurately represented in COMSOL, as it is imported from the CAD designs used for machining of the HFNG's structural components. The assumptions built into the simulations are as follows:

1) Ions are extracted from a flat plasma meniscus, which is a reasonably good assumption because of the small size of the extraction aperture and when operated near the Child-Langmuir limit \cite{Plasma}.

2) All ions are born with the same speed and with a perpendicular velocity to the surface of the plasma. Even though the energy of the ions in the plasma follows a Maxwell-Boltzmann distribution, this assumption encompasses the overall behavior of the beam as it travels to the target. 

\subsection{Single-Hole Extraction System} \label{section:single extraction}

\begin{figure}
	\centering
	\subfloat[\label{subfig:beam_spread}]{\includegraphics[width=0.55\textwidth]{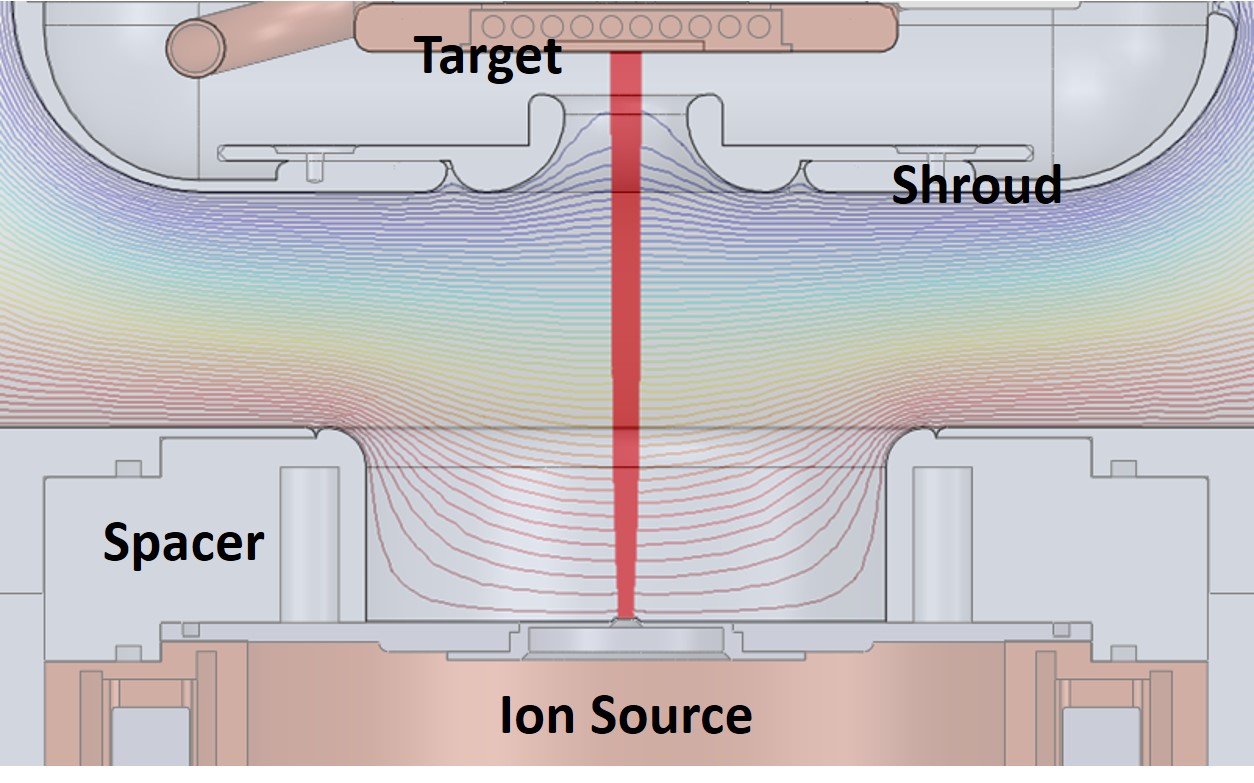}}
	\hfill
	\subfloat[\label{subfig:beam_in_shroud}]{\includegraphics[width=0.445\textwidth]{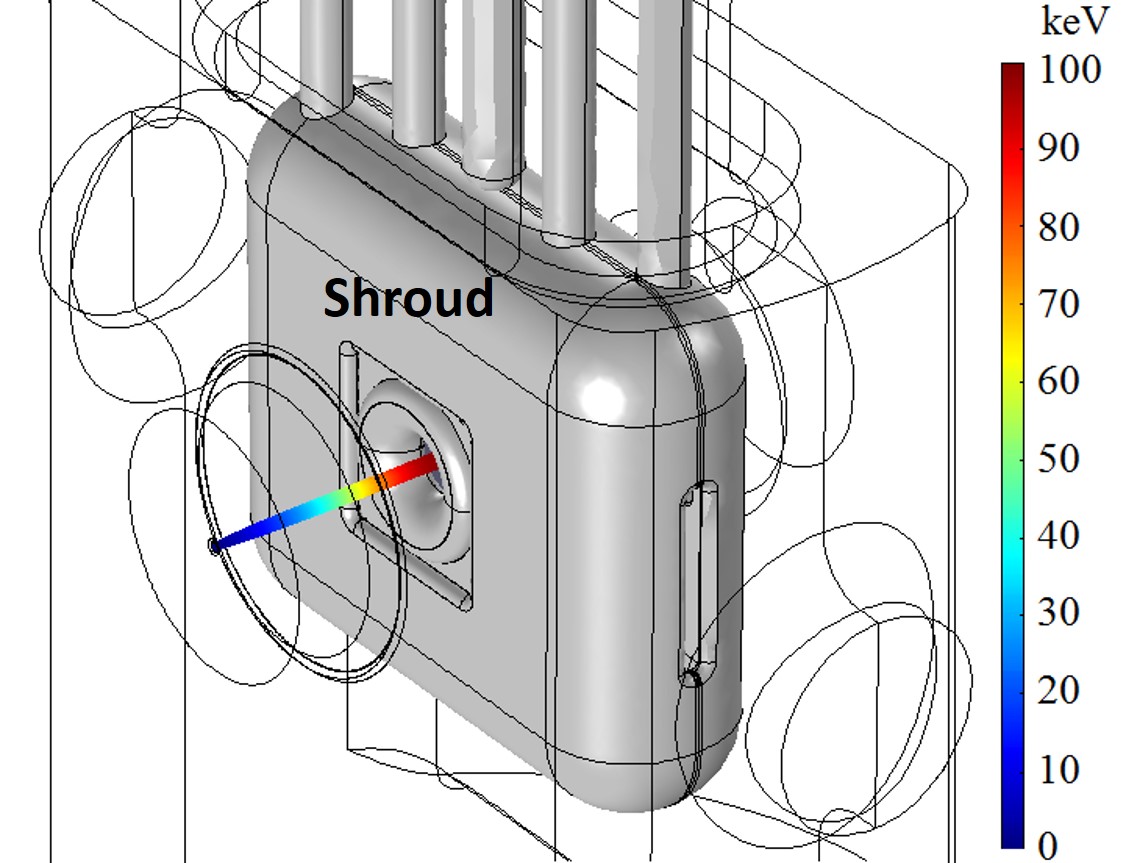}}
	\caption{Comsol simulation of electric field and deuteron beam trajectory at 100 kV and 1.4 mA (single-aperture extraction). Note the focusing electric field due to the spacer.}
	\label{fig:beam_spread}
\end{figure}

Figure \ref{fig:beam_spread} shows the electric potential and the simulated particle trajectories at a target bias and beam current of 100 kV and 1.4 mA, respectively. Note in Figure \ref{subfig:beam_spread} that the equipotential lines near extraction result in a slightly focusing field. This is due to the HFNG geometry which requires that the ion source be recessed back a certain distance in order to achieve the desired current density, reduce the maximum electric field in that region, and allow the beam to spread so the heat flux on the surface becomes acceptable. This was achieved with an aluminum spacer of 3 cm, as shown in Figure \ref{subfig:beam_spread} 

\begin{figure}
	\centering
	\includegraphics[width=0.7\textwidth]{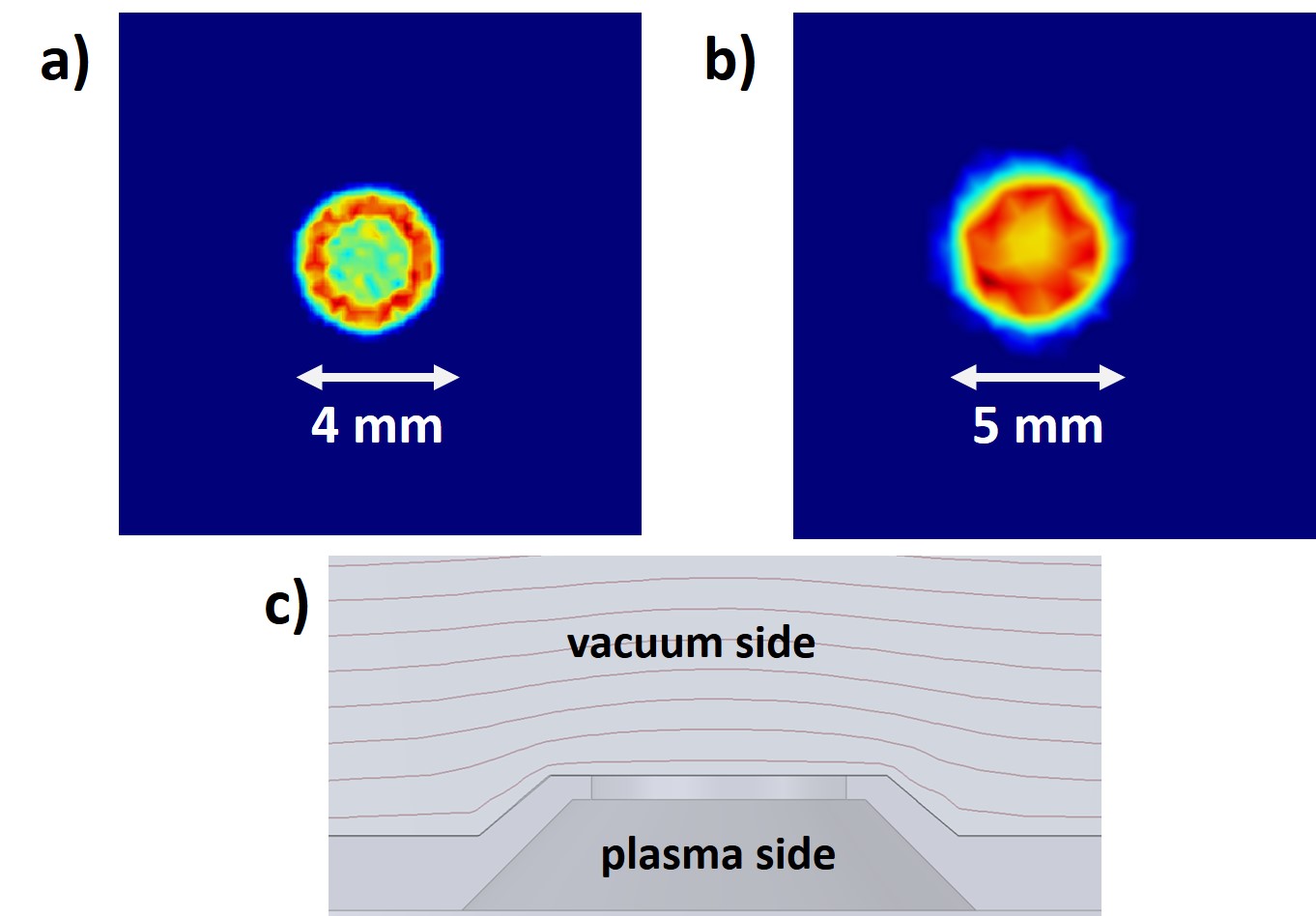}
	\caption{Beam spreading by modification of the plasma electrode: a) flat extraction, b) \& c) convex extraction. The nozzle rises 0.635 mm above the base of the plasma electrode, and has a chamfer angle of 45$^\circ$.}
	\label{fig:beam_spot3}
\end{figure}

The resulting beam spot size and heat flux on the target are shown in Figure \ref{fig:beam_spot3}. Note that the beam profile in Figure \ref{fig:beam_spot3}a is non-uniform in the radial direction, and localized heating is observed around the edges. Therefore, a convex-shaped extraction nozzle was designed, whose purpose is to counteract the focusing electric field experienced near the aperture. The nozzle design is shown in Figure \ref{fig:beam_spot3}c, and further details about it are described in \cite{CoryThesis}. Note the defocusing effect of the nozzle due to the fact that the equipotential lines are slightly convex downstream of the beam path. The resulting beam spot is not only more uniform, but it is also spread into a larger area, which reduces the power density on the target.

\subsection{Multiple-Hole Extraction System} \label{section:mult beam extraction}

\begin{figure}
	\centering
	\includegraphics[width=0.6\textwidth]{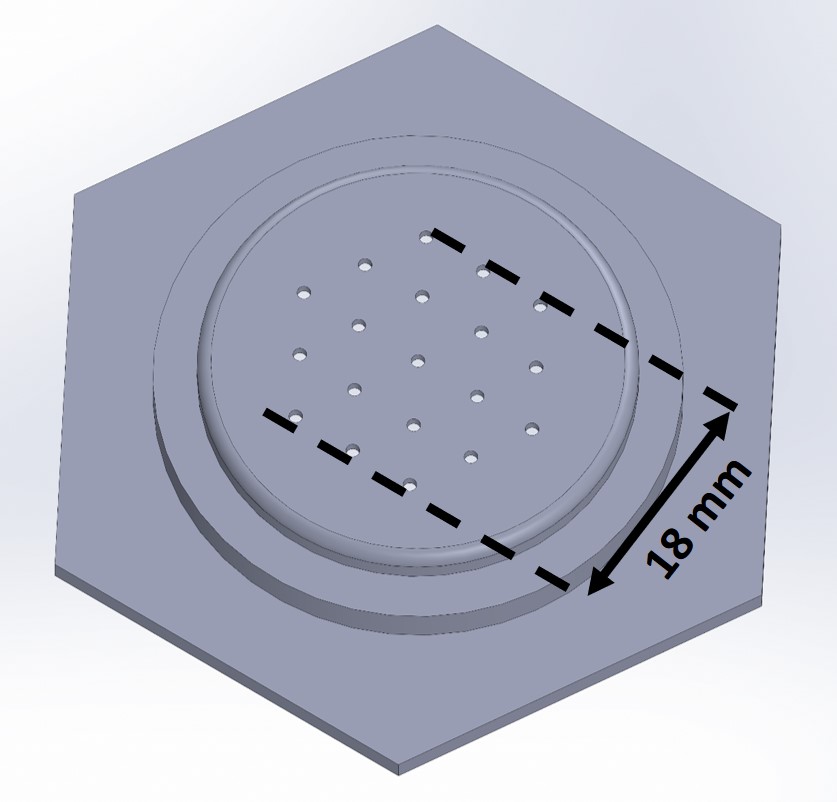}
	\caption{CAD drawing of an optimized 19-hole plasma electrode with a hexagonal pattern. Each individual aperture is 1 mm in diameter and the top is not flush against the interior of the vacuum chamber, but rather protrudes 1 mm in order to shape the electric field around it (convex equipotential lines). }
	\label{fig:CAD_extraction}
\end{figure}

The uniformity of the beam spot and further spread of impingement area can be achieved by alternative means other than optimizing a single extraction nozzle. For instance, an Einzel lens configuration, \emph{i.e.,} one or more extraction plates located downstream of the beam and biased at different potentials can give further control of the beam envelope. One of the downsides of such an arrangement is the higher degree of complexity added to the design, which stems from biasing, insulating, and properly installing these plates inside the vacuum chamber. Another option is the use of multiple apertures in the plasma electrode. Figure \ref{fig:CAD_extraction} shows an optimized design of a 19-aperture plasma electrode arranged in a hexagonal pattern, which was chosen in order to optimize the packing fraction and the uniformity of the beam. The packing fraction is essential for maximizing the neutron flux at the sample location. Additionally, the top of the plasma electrode protrudes out a few millimeters inside the chamber in order to achieve an initial defocussing of the deuterium beam, as explained in \ref{section:single extraction}.  

In order to approximate the maximum beam current that can be extracted from each individual hole, the Child-Langmuir law was employed (Equation \ref{eqn:child-lang}), where $j_c$ is the ion current density \emph{i.e.,} extractable current divided by the area of the aperture, $\epsilon_0$ is the permittivity of free space, $Ze$ is the extracted ion charge, $m_i$ is the ion mass, $V_0$ is the extraction potential, and $d$ is the distance between the plasma electrode and the extraction electrode, which is the shroud in the case of the HFNG ($d=$ 7 cm). This equation sets a limit for the extractable current through an aperture assuming the system is space charge-limited (enough ions are available in the ion source). This treatment was derived only for beam extraction between parallel plates, but serves as a good approximation for a variety of distinct geometries \cite{Plasma}. 

\begin{equation}
j_c=\frac{4}{9}\epsilon_0 \sqrt{\frac{2Ze}{m_i}}\frac{V_0^{3/2}}{d^2}
\label{eqn:child-lang}
\end{equation}

\begin{figure}
 	\centering
 	\includegraphics[width=0.8\textwidth]{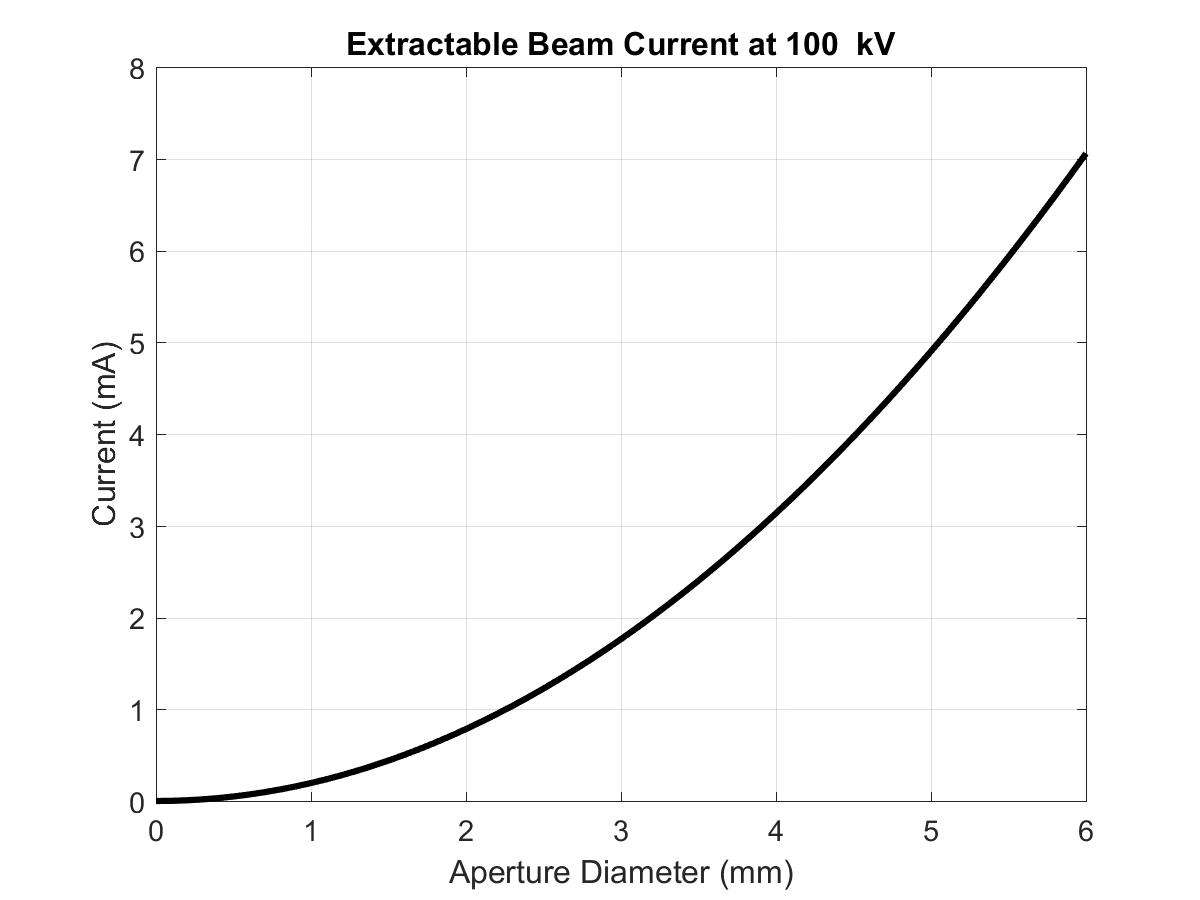}
 	\caption{Maximum extractable current according to Child Langmuir law for the HFNG geometry at 100 kV \cite{Humphries2}.}
 	\label{fig:child_lang}
 \end{figure}

Figure \ref{fig:child_lang} shows the maximum beam current that can be extracted as a function of the aperture diameter for the HFNG geometry at 100 kV. Because it is desirable to keep the diameter of the apertures small in order to avoid plasma meniscus issues, as explained above, we chose a diameter of $1$ mm, through which it is possible to draw a current of $\approx 0.196$ mA. Therefore, for the 19-hole plate arrangement, it is theoretically possible to extract up to $3.7$ mA of beam current, or a total of $7.4$ mA if both ion sources are used. This value for the current is an approximation, but it serves as a conservative input parameter for ion beam optics simulations. In fact, the observed current limit was 3.5 mA for a single ion source, confirming the validity of this approximation.

Simulations show a more uniform beam profile near the center of the target and a larger impingement area, which translates into a lower heat flux than that of a single aperture even at higher currents. Beam current can be further increased by adding more apertures or by increasing the diameter of each individual aperture. The increase in current is limited by the maximum number of apertures that can be drilled on the plate or by the increase in chamber pressure due to a larger open area between the ion source and the rest of the vacuum chamber. The latter could make the gas pressure too high for optimum operation. Several operating pressure values were tested (up to $1\times 10^{-4}$ mTorr) without significant increase in beam loss or arcing frequency. Ultimately, the beam current is limited by the properties of the ion source and available RF power. 

\begin{figure}
	\centering
	\includegraphics[height=0.6\textwidth]{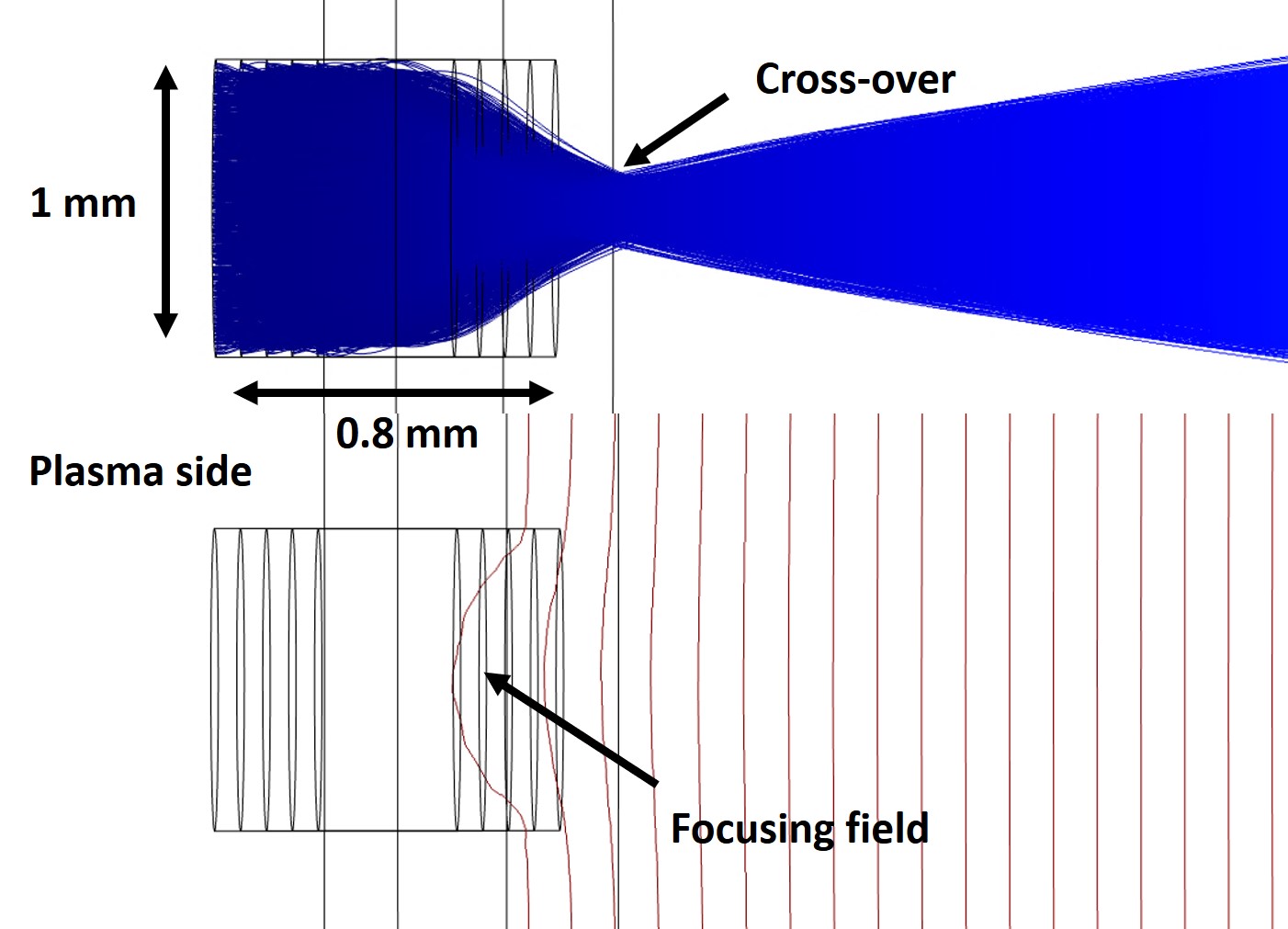}
	\caption{Simulation of deuteron trajectories near extraction (one aperture). Deuterium ions are assumed to be extracted at a few electron-volts and perpendicular to the surface of a flat plasma meniscus. Note the convergent field (b) that provides the initial focusing force that is ultimately responsible for the spreading of the beam.}
	\label{fig:cross_over}
\end{figure}

\begin{figure}
	\centering
	\includegraphics[height=0.46\textwidth]{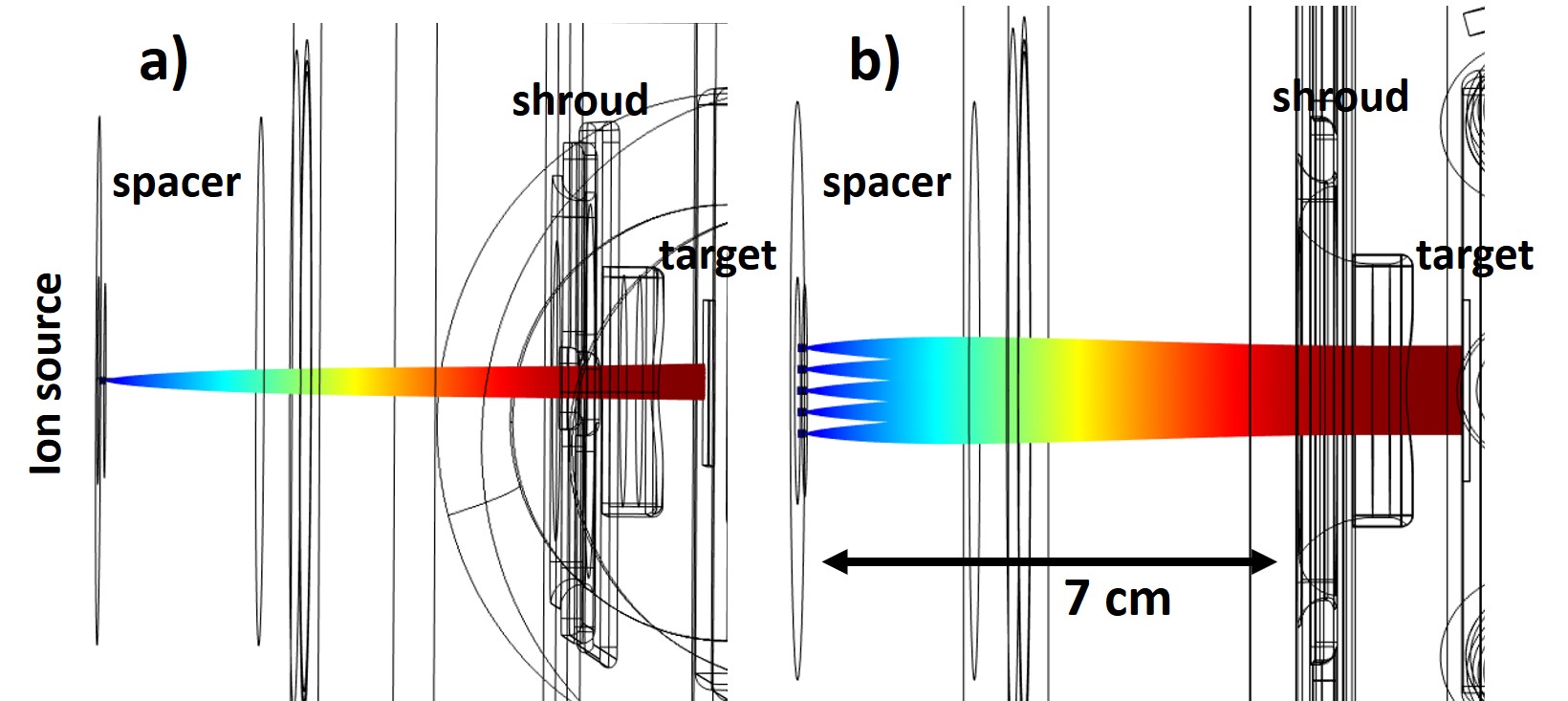}
	\caption{The beam envelope for a single beam, a), located at the center is divergent while that of multiple apertures, b), is convergent because of the addition of the spacer.}
	\label{fig:single_vs_mult}
\end{figure}

The geometry of the HFNG is such that there exists a focusing electric field at extraction, as shown in Figure \ref{fig:beam_spread}. Therefore, off-centered beamlets experience a focusing force, which depends on their location with respect to the center of the plasma electrode. Beamlets further away from the center experience a stronger focusing force because the equipotential lines are more concave in that region. Therefore, multiple-extraction aperture systems produce a convergent beam envelope. In contrast, a single hole in the center produces a divergent beam envelope, as shown in Figure \ref{fig:single_vs_mult}. Regardless of the type of plasma electrode, each individual beamlet spreads out as it travels along the acceleration gap, resulting in a beam spot larger than the diameter of the aperture (a factor of 3 for a 1 mm aperture in the case of the HFNG). The reason for this spread is due to the initial focusing provided by the extraction aperture itself, which causes deuterium ions to cross over and then diverge. This phenomenon is depicted in Figure \ref{fig:cross_over}. Essentially, each extracting aperture acts as a non-linear electrostatic lens with spherical aberration (not focused to a point) \cite{Plasma}. 

\begin{figure}
	\centering
	\includegraphics[width=0.5\textwidth]{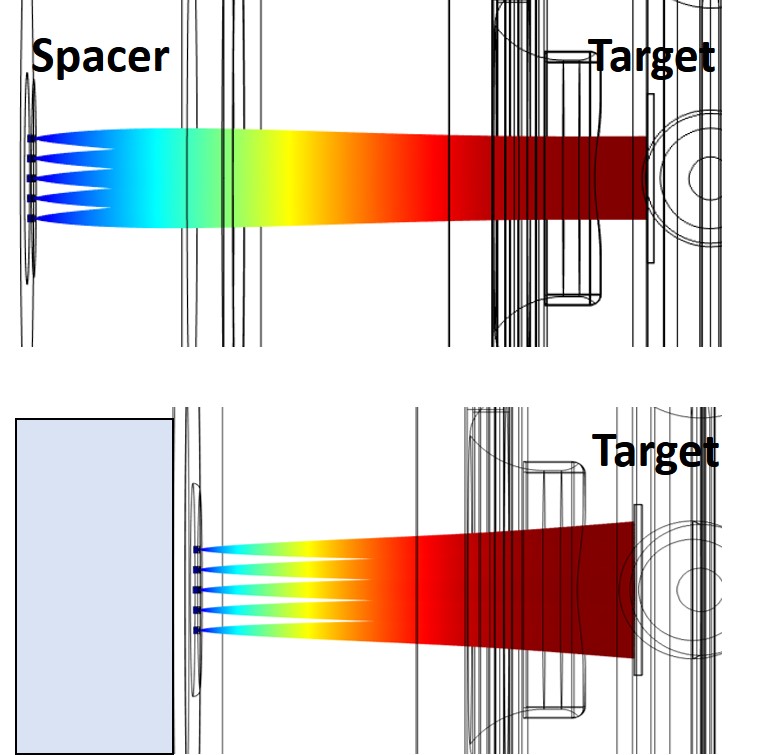}
	\caption{Beam envelope with spacer (top), and without spacer (bottom), showing the effect of the spacer, which gives an additional standoff of 3 cm; the spacer also introduces a focusing field.}
	\label{fig:spacer_vs_noSpacer}
\end{figure}

If there were no spacer, the off-centered beamlets would not experience an overall focusing force and the beam envelope would be divergent, as shown in Figure \ref{fig:spacer_vs_noSpacer}. However, at such a close distance between the shroud and the plasma electrode ($\approx 7$ cm), the individual beamlets would not spread enough and the heat deposition per beamlet would be unacceptable.  

\begin{figure}
	\centering
	\includegraphics[height=0.7\textwidth]{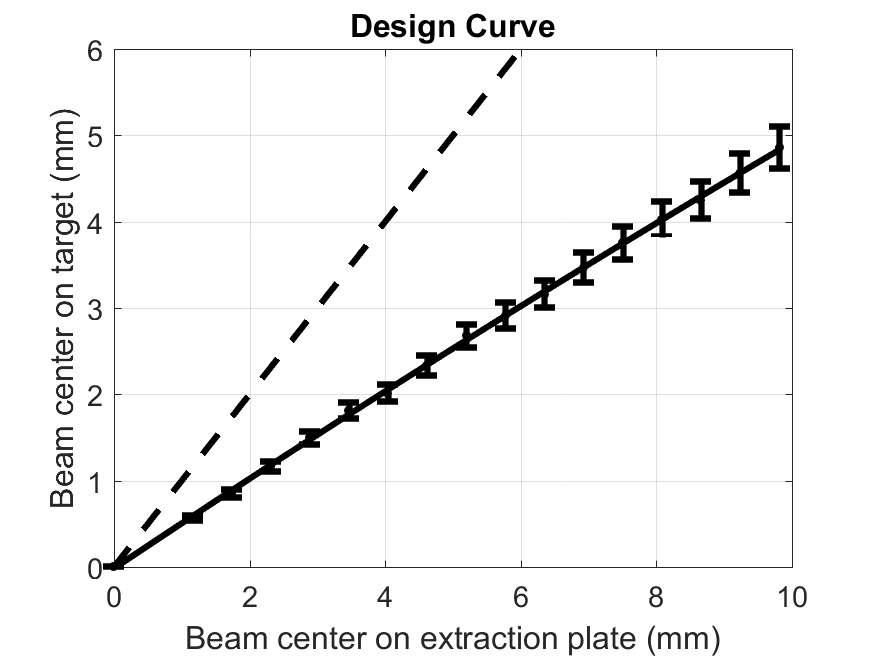}
	\caption{Design curve used to estimate the center of the individual deuterium beamlets on the target. The dashed curve is for the case with no spacer (linear) and the solid curve is the result of COMSOL simulations for the current HFNG configuration.}
	\label{fig:designC}
\end{figure}

Simulations were performed in order to optimize the hole pattern distribution on the plasma electrode. Because of the geometrical symmetry, only one beamlet was simulated by varying the position of the 1 mm-diameter aperture relative to the center of the plasma electrode in steps of 1 mm (up to 10 mm). The location of the apertures were chosen so that the heat flux on the target is as uniform as possible. The design curve shown in Figure \ref{fig:designC} shows the results of such simulations. The y-axis shows the beamlet center on the surface of the target as a function of the aperture position on the plasma electrode. The origin represents the center of the plasma electrode and the center of the target, respectively. The dashed line shows the hypothetical case in which there is no focusing field (no spacer) and the beamlets are extracted horizontally. In this case, the relationship between these two parameters is one-to-one \emph{i.e.,} an aperture located 1 mm away from the center of the plasma electrode would result in a beam spot whose center is exactly 1 mm away from the center of the target. Note that in either case, each beamlet spreads out as they travel along the acceleration gap. 

The optimization process focused on two parameters: reduction of heat flux on the target surface, and uniformity of the heat flux distribution, which translates into a more uniform neutron flux distribution at the sample location.

\begin{figure}
	\centering
	\includegraphics[height=0.8\textwidth]{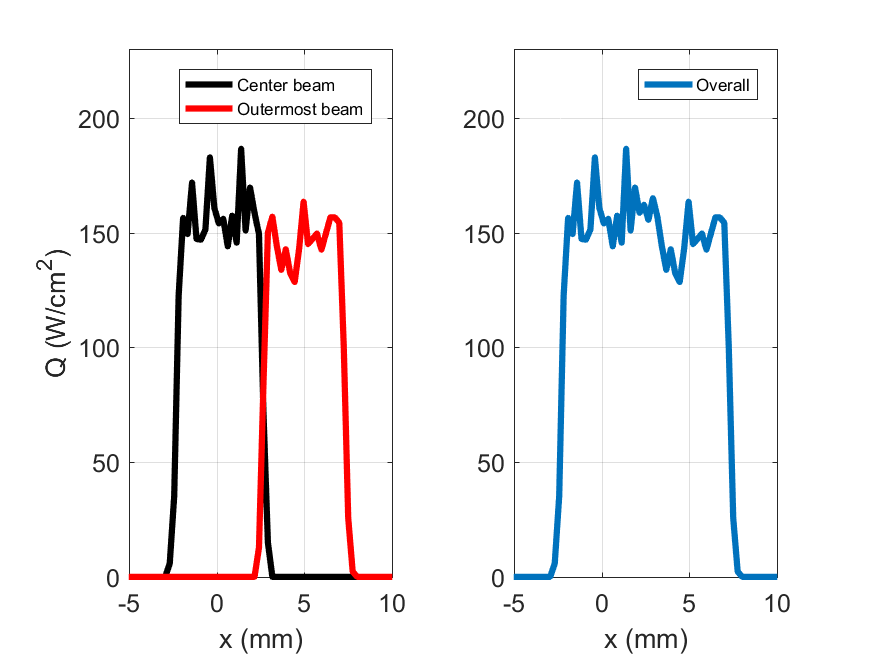}
	\caption{Comsol simulation results showing the heat flux on the target due to the center and outermost beamlets. The distance from the center of the plasma electrode of the outermost aperture latter was chosen in order to achieve a flat profile.}
	\label{fig:two_beams}
\end{figure}

\begin{figure}
	\centering
	\includegraphics[height=0.8\textwidth]{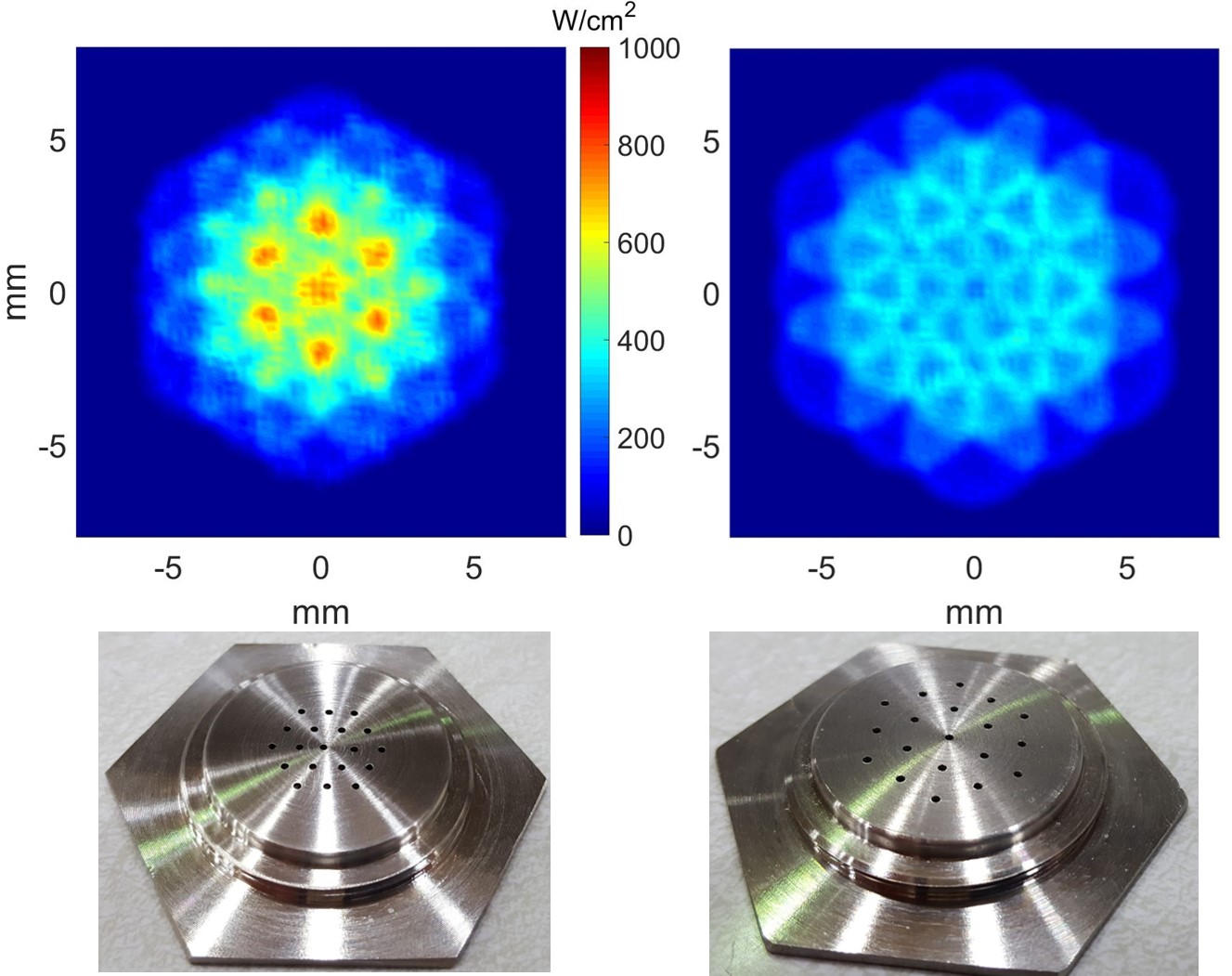}
	\caption{Heat flux on the target for a non-optimized design (left) and a design optimized according to the design curve shown in Figure \ref{fig:designC} (right). The overall size of the beam on target does not change significantly, but the peak heat flux is reduced from approximately 1 kW/cm$^2$ to less than 600 W/cm$^2$. The diameter of the apertures are 1 mm each and they are arranged in a hexagonal pattern to optimize the packing fraction.  }
	\label{fig:heat maps}
\end{figure}

Figure \ref{fig:two_beams} shows the simulated 2D beam profiles of the center beamlet located at (0,0) and the outermost beamlet. The latter was placed 9.8 mm from the center of the plasma electrode in order to achieve a uniform beam profile on the target. The intermediate hole was placed in such a way that its corresponding beamlet center would hit the target exactly midway between the other two. This location (center-to-center) was calculated (from the design curve) to be 5.1 mm from the center. The resulting heat flux distribution (W/cm$^2$) is shown in Figure \ref{fig:heat maps} (upper right). Note that evenly spaced extraction apertures (on the plasma electrode) result in a non-uniform distribution on the target due to the uneven focusing effect explained previously. A design leading to an unacceptable heat profile is shown in Figure \ref{fig:heat maps} (left). 

Note that even though the spacing between the apertures is larger on the optimized plasma electrode design ($d \approx 5$ mm), the resulting overall beam profile on the target is comparable in size as the non-optimized design ($d\approx 3.5$ mm), but with a higher degree of uniformity and reduced heat flux. Experimental tests validate these simulations, as shown in the experimental validation section.

\subsection{Heat Transfer Analysis}

\begin{figure}
	\centering
	\includegraphics[height=0.5\textwidth]{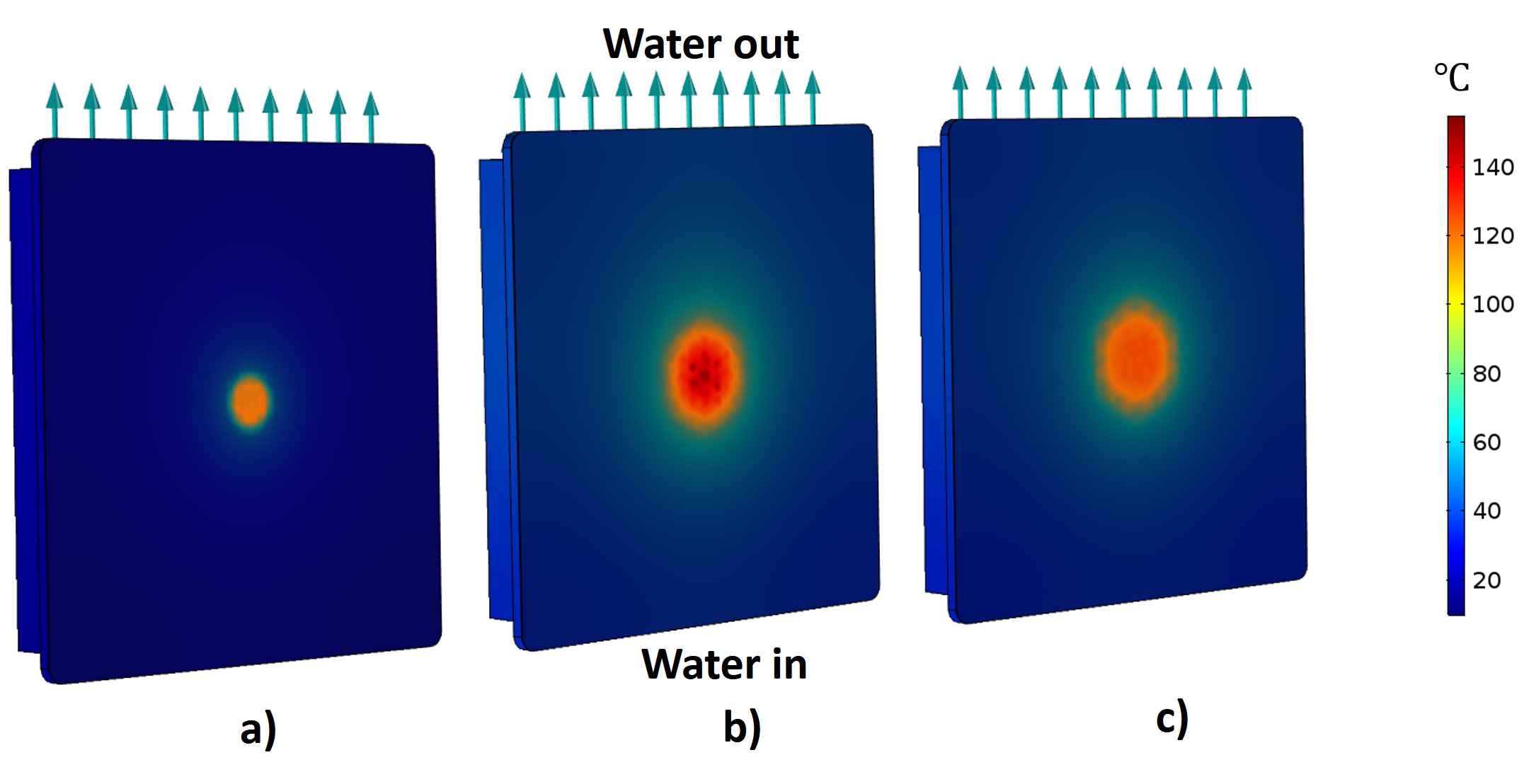}
	\caption{Temperature maps on the target for different plasma electrodes a) single-hole b) multiple-apertures, non-optimized, and c) Multiple-apertures optimized. Maximum temperature values are $104^{\circ}C$,  $155^{\circ}C$,  $114^{\circ}C$, respectively}
	\label{fig:temp_comsol}
\end{figure}

The heat flux distribution on the target surface was modeled in COMSOL using the heat transfer in solids module coupled with the turbulent flow module. Further details about the target's heat transfer analysis may be found in \cite{CoryThesis}. Figure \ref{fig:temp_comsol} shows the temperature distribution on the surface of the target for three different beam profiles: single hole ($d=$ 0.262 cm), non-optimized multiple-hole (pitch = 3.4 mm, $d=$ 1 mm), and optimized multiple-hole plasma electrode (pitch = 5 mm, $d=$ 1 mm). Note that all plasma electrode designs show maximum temperature values below the degassing temperature of hydrogen in titanium ($\approx 200\ ^{\circ}C)$ \cite{CRC}. However, the simulations do not account for the inter-metallic phases between the copper and the titanium, or the bonding process itself, which are extremely important for accurate modeling of the heat transfer between the titanium layer and the copper piece. However, the main purpose of these simulations is not to accurately predict the surface temperature of the target, but rather to be able to compare the relative heat transfer effectiveness of future target designs. Additionally, variations in titanium thickness and the purity of the metal affect the neutron yield; thinner titanium layers have better heat transfer properties in the target. The disadvantage of having a very thin titanium layer (a few microns thick) is that the target has a shorter lifetime due to ion sputtering.

Experimentally, it was determined that there is some overheating of the target in the case of the non-optimized plasma electrode. This was evidenced by the initial increase in neutron dose rate rapidly followed by a decrease in dose rate as measured by a Bonner sphere. The neutron yield in the case of the optimized plasma electrode allows for an increase in neutron yield by a factor of $\approx 2.6$ as expected from Equation \ref{eq:yield}, which shows a linear relation between ion beam current (proportional to $\phi_d$) and neutron yield.

\section{Experimental Validation of the Neutron Flux and Deuteron Beam Profile} \label{section:exp_valid}	

The neutron flux was experimentally determined through neutron irradiation of a single indium foil located at the center of the sample holder for both a single-aperture and an optimized multiple-aperture plasma electrode. Indium is a soft metal which can be easily cut into foils of approximately 0.9 cm in diameter and 1 mm thick. The naturally occurring isotope $^{115}$In has a long-lived ($^{115m}$In, $t_{1/2}=4.486 \pm 0.004$ h, IT = 95.0 $\pm 0.7$\% to $^{115}$In) \cite{NDS} isomer which decays by the emission of a 336.241 keV gamma ray to its ground state; this isomer is populated by the inelastic scattering reaction $^{115}$In(n,n$^{\prime}$)$^{115m}$In. Since this reaction is a threshold reaction, low-energy neutrons have a very low probability of populating this excited state. Moreover, the cross-section does not present a large gradient within the energy window of interest (2.1-2.8 Mev) \cite{NDS}, which means that it is not necessary to integrate the cross-section over the angle subtended by the foil as long as its diameter is relatively small. This factors out the potential errors in the geometry of the foil. Additionally, the reaction $^{115}$In(n,$\gamma$)$^{116}$In also leads to a strong gamma signal. Therefore, both reactions provide insight into two different parts of the spectrum. Other foils used as flux monitors such as gold or Dysprosium are usually used in reactors because of the large capture cross-section of Au-197 and Dysprosium at low neutron energies, and because of this, they are not ideal for our application. Through the combination of MCNP modeling of neutron transport and decay spectroscopy of the activated indium foil, the energy window and flux distribution subtended by the foil can be well-characterized for a given target configuration. Further details on the flux characterization can be found in the recent work of A.S. Voyles \emph{et al.} \cite{np_paper}.

Table \ref{tab:flux} shows the measured neutron flux after approximately 3 hours of irradiation for the experimental arrangements explained above.

\begin{table}
	\centering
	\caption{Flux analysis: comparison between different designs. Both measurements were performed with the same indium foil located at the center of the HFNG sample holder.} 
	\label{tab:flux}
	\begin{tabular}{ |P{5cm}|P{5cm}| }
		\hline
		\textbf{plasma electrode design}   & \textbf{Average flux (n/cm$^2$/s) 8 mm  away}   \\
		\hline
		Single aperture (D = 0.262 cm)          &  1.04$\times$10$^7$ $\pm$ 7\%          \\
		19-apertures (D$_i$ = 1 mm)             &  2.30$\times$10$^7$ $\pm$ 5\%         \\
		\hline
	\end{tabular}
\end{table}

Based on the modeling detailed in section \ref{section:mult beam extraction}, it is expected an increase in neutron yield of a factor of 2.6 (i.e. 3.7 mA/1.4 mA). The increase in flux was measured to be a factor of approximately 2.2. It is expected that the flux increases at a lower rate (for a given sample size) than the neutron yield due to the increase in the diameter of the beam on the surface of the target.   

\begin{figure}
	\centering
	\includegraphics[width=1\textwidth]{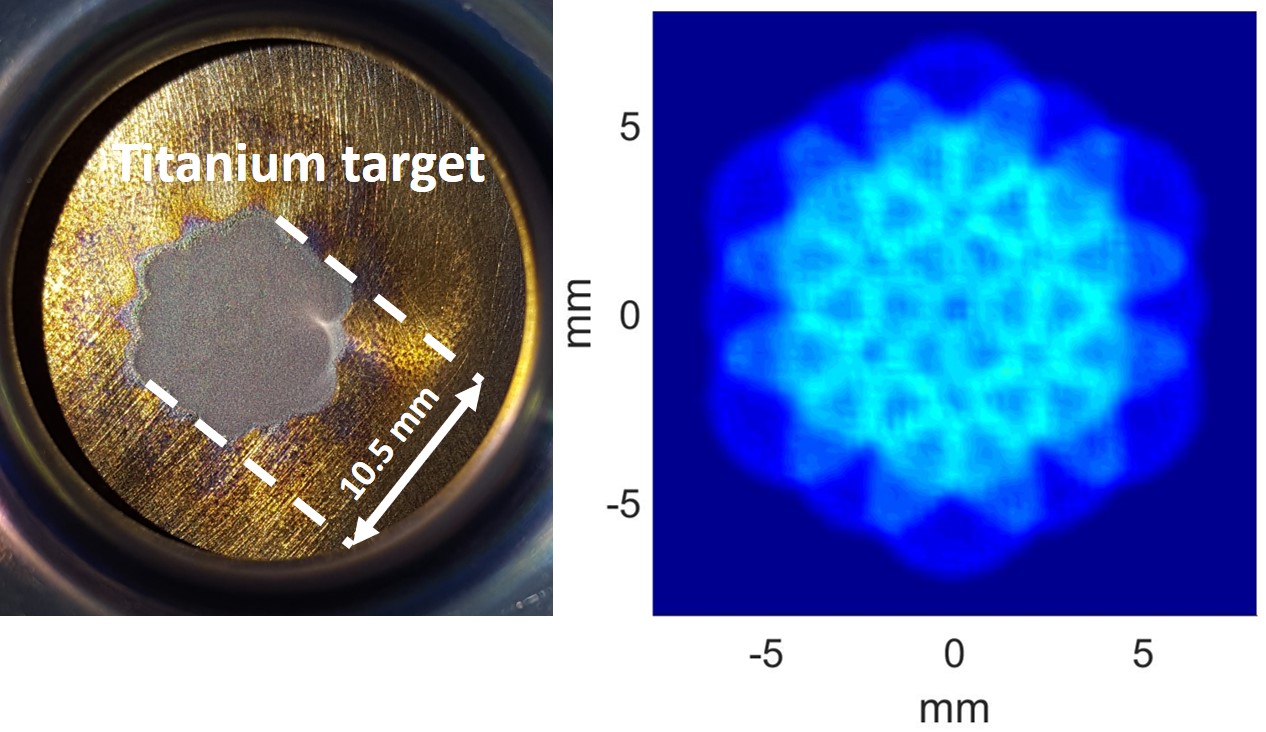}
	\caption{Optimized 19-aperture plasma electrode design and resulting beam spot. Overall size and uniformity agree well with simulations (right).}
	\label{fig:experiment_beam_prof}
\end{figure}

Figure \ref{fig:experiment_beam_prof} (left) shows the beam profile on the titanium target resulting from the multi-aperture plasma electrode after several hours of irradiation. The simulated beam profile is also shown in Figure \ref{fig:experiment_beam_prof} (right). Note the remarkable agreement between the actual and the predicted beam profile.    

A version of this optimized multiple-aperture plasma electrode was used to irradiate a geological sample over a period of several weeks.

As the deuterium beam hits the target, it forms titanium deuteride with varying amounts of deuterium dissolved in the matrix. It was observed that after a few minutes of irradiation, the beam spot is colored black, but after several hours of irradiation, the beam spot turns light gray, which might be an indication of the amount of deuterium absorbed in the target.

\section{Conclusions and Outlook}

The HFNG is a multi-purpose, versatile neutron generator which has been well-characterized in terms of flux and energy distribution. Reliable simulation tools have been developed and benchmarked through experimental validation of the generator operation. These tools include modeling of the HFNG beam optics, heat transfer, neutron flux, and neutron energy distribution. The neutron flux has been experimentally determined via indium activation foils. Experimental validation agrees with the simulations described here to within approximately 5\% at the sample holder location.

Moreover, neutron reflection studies are being investigated in order to increase the neutron fluence in the sample location. Dense compounds such as lead or calcium carbonate act as efficient neutron reflectors, minimizing the neutron energy loss per collision, which prevents significant softening of the neutron spectrum.

The heat removal capability of the target is one of the most important limiting parameters for increasing the generator's current neutron flux since deuterium degases from the titanium target without sufficient heat transfer. Moreover, the beam spot size must remain small in order to maximize the flux at the center of the sample holder. While it is possible to increase the yield by intentionally spreading the beam spot, this would correspond to a non-linear increase in flux, which complicates modeling of higher current designs.

As the beam spot size increases, suppression of secondary electrons also becomes a significant challenge. The opening of the shroud (suppression electrode) becomes too large to establish an electric field in the center capable of suppressing secondary electrons, hence different techniques need to be pursued.

Finally, there is a lack of peer-reviewed literature regarding different target materials other than titanium that can serve as efficient deuterium getters and at the same time have outstanding heat transfer properties. 

The potential of high-flux neutron generators for radioisotope production cannot be overestimated since that could change the landscape of medical isotope production and utilization.   

\section{Acknowledgments}

We gratefully acknowledge a grant from the University of California Office of the President.
	
Work supported by NSF Grant No. EAR-0960138, U.S. DOE LBNL Contract No. DE- AC02-05CH11231, and U.S. DOE LLNL Contract No. DE-AC52-07NA27344.

D.R. was supported by DFG research scholarship RU 2065/1-1.
	
\section*{References}
	
\bibliography{mybibfile}
	
\end{document}